\documentclass[12pt]{iopart}
\usepackage{iopams}
\usepackage{graphicx}
\eqnobysec
\newcommand{\cm}{\mbox{cm}}
\newcommand{\be}{\begin{eqnarray}}
\newcommand{\ee}{\end{eqnarray}}

\def\ben{\begin{equation}}
\def\een{\end{equation}}
\def\bena{\begin{eqnarray}}
\def\eena{\end{eqnarray}}

\renewcommand{\d}{\mbox{{\rm d}}}

\begin{document}

\begin{flushright}
DAMTP-2006-101\\ SISSA 70/2006/A
\end{flushright}

\title{The Cosmological Slingshot Scenario: A Stringy Early Times Universe}

\author{Cristiano Germani}
\address{D.A.M.T.P., Centre
for Mathematical Sciences, University of Cambridge, Wilberforce
road, Cambridge CB3 0WA, England \\ Department of Physics, King's College London, London WC2R 2LS, England\\
SISSA, via Beirut 4, 34014 Trieste,
Italy} \ead{Germani@sissa.it}

\author{Nicol\'as Grandi}
\address{IFLP-CONICET, cc67, CP1900 La Plata, Argentina\\
ICTP, Strada Costiera 11, 34014 Trieste, Italy\\
SISSA, via Beirut 4, 34014 Trieste, Italy}
\ead{grandi@fisica.unlp.edu.ar}

\author{Alex Kehagias}
\address{Physics Division, National Technical University of Athens,\\ 15780 Zografou Campus,  Athens, Greece}
\ead{kehagias@central.ntua.gr}
\begin{abstract}
A cosmological model for the early time Universe is proposed. In this model, the Universe
is a wandering brane moving in a warped throat of a Calabi-Yau
space. A
non-zero angular momentum induces a turning point in the brane
trajectory, and leads to a bouncing cosmology as experienced by an
observer living on the brane. The Universe undergoes a decelerated
contraction followed by an accelerating expansion and no big-bang
singularity. Although the number of e-folds of accelerated motion is low (less than 2),
standard cosmological problems are not present in our model thanks
to the absence of an initial singularity and the violation of
energy conditions of mirage matter at high energies. Density
perturbations are also calculated in our model and we find a
slightly red spectral index with negligible tensorial
perturbations in compatibility with WMAP data.
\end{abstract}

\maketitle

\section{Introduction}
In April 1970 the spacecraft Apollo 13 was sent to the moon for a
NASA mission. Two days after its launch an explosion made the
Service Module lose power and oxygen very  quickly. In this
dangerous scenario the astronauts had to take the decision to move
into the Lunar Module {\em Aquarius} and leave the Service Module.
However, as the Lunar module is not designed for long trips, NASA
engineering had to find a way to bring back to earth the Lunar
Module, using as less energy as possible. As the Module was moving
towards the Moon, NASA rocket scientists decided to use very
little power of the Module in order to modify its orbit to be an
open orbit around the Moon. This was lead by the knowledge that,
at the inversion point of the trajectory of the (probe) spacecraft
around the Moon, conservation of angular momentum would lead to an
acceleration of the spacecraft and would redirect it towards the
Earth. The effect they used is called the gravitational Slingshot
effect.

In this paper we show that the same effect appears any time a
brane moves with an open orbit around a not trivial central
background. In this case an observer living on the brane will
experience a bouncing cosmology without singularity with a well
defined bounce, followed by a very short acceleration period. We
will call this the ``Cosmological Slingshot'' scenario.

We will study the cosmological evolution of a brane observer
at early times using the so called mirage cosmology approximation
\cite{Kehagias:1999vr},\cite{Kiritsis:2003mc}. In it, what the observer is
measuring as the physical $3+1$ dimensional metric is given in
terms of the brane embedding and the bulk metric by the induced
metric formula. Its time evolution is then dictated by the motion
of the brane in the background which, if the back-reaction of the
probe can be neglected, is a static solution of $9+1$ dimensional
IIB supergravity equations.
In particular, our Slingshot brane is moving on a Calabi-Yau manifold sourced by a stack of $D3$-branes.
In this background, mirage effects can dominate the evolution of the Universe only at
early time, {\it i.e.} when the brane moves in the throat of the Calabi-Yau. Viceversa, at
late times, when the brane reaches the hat of the Calabi-Yau, local effects {\it a la} Randall-Sundrum \cite{Randall:1999vf} might become important thus
obtaining the standard cosmological evolution at late times.
In this context, we will show that there is an infinitely large region of choice
of parameters such that homogeneity, isotropy and flatness
problems of standard cosmology can be avoided as follow:
\vspace{.1cm}

{\noindent\bf{Homogeneity}} Latest measures of the cosmic microwave
background (CMB) \cite{Spergel:2006hy} clearly show that the CMB radiation
is homogeneous and isotropic. This implies that all points of the
CMB sky, as we observe them nowadays, should have been in causal
contact, with high degree of accuracy, at very early times. In
other words, two photons from two distant points in the CMB sky
should have had enough time in the past to meet each other and
exchange information. Standard Big Bang cosmology predicts a
finite ``life time'' of the Universe. One can show that this
``life time'' is too short to allow the high homogeneity and
isotropy of the CMB sky.

The Slingshot scenario overcomes this problem by generating a
non-singular bouncing cosmology. In this case two photons had
enough time to meet each other in the past so that the high degree
of homogeneity and isotropy observed could be reached. More
technically, the conformal horizon of the Universe in the
Slingshot scenario is infinite.
\vspace{.1cm}

{\noindent \bf{Isotropy}} Standard Cosmology assumes that the cosmic
evolution is driven by ``standard matter'' (which does not violate
energy conditions) via Einstein equations. It is possible to prove
that in this case an isotropic Universe is highly unstable under
anisotropic perturbations. This is because, close to the
singularity, the energy density associated with anisotropic
perturbations is stronger that the energy density of standard
matter.

In the Slingshot scenario, this problem is solved in two different
ways. Firstly, our bouncing cosmology may not allow the Universe
to reach scales in which the anisotropic perturbations dominates.
Secondly, one can show that for small scales, the mirage energy
density grows faster than the energy density associated with
anisotropic perturbations, so to stabilize an homogeneous and
isotropic Universe, a Friedman geometry.
\vspace{.1cm}

{\noindent \bf{Flatness}} Current measures \cite{Spergel:2006hy}, clearly show
that our Universe is nowadays very close to be spatially flat.
However, in standard cosmology a measure of the spatial curvature
decreases backwards in time while the Hubble constant increases.
In this case in order to have the required flatness today, extreme
fine tuning of the curvature must be used at very early times
close to the Planck scales. This is mainly due to the fact that,
given the matter content, spatial curvature depends only on one
parameter.

In the Slingshot scenario we solve this problem as
the spatial curvature is bounded from below. Indeed, by an appropriate choice of
the minimum for the spatial curvature,
which only restrict to a semi-infinite region the space of parameter of our
model (the ``energy'' and
``angular momentum''), the flatness problem is easily
solved without the use of fine-tunings.
\vspace{.1cm}

Besides the resolution of standard cosmological problems, we show
that the correct scale invariant perturbation necessary to
reproduce the CMB spectra as measured by the WMAP team
\cite{Spergel:2006hy}, can be easily produced. This is due to the fact that
perturbations are dynamically generated at a stringy scale. The
introduction of this scale is indeed enough to produce an almost
scale invariant spectrum.

\subsection{Framework}

The bosonic part of the bulk supergravity action (in Einstein frame) for the type IIB
theory is \cite{Schwarz:1983wa}
\begin{eqnarray}
\label{action0}
S_{IIB}&=&\frac{1}{2\kappa_{10}^2}\int d^{10}x\sqrt{-g}
\left(R-\frac{\partial_M\tau\partial^M\bar\tau}{2(\mbox{Im}\tau)^2}-\frac{G_{(3)}\cdot \bar G_{(3)}}
{12\mbox{Im}\tau}-\frac{\tilde F_{(5)}^2}{4\cdot 5!}\right)\cr
&+&\frac{1}{8i\kappa_{10}^2}\int \frac{C_{(4)}\wedge G_{(3)}\wedge\bar G_{(3)}}{\mbox{Im}\tau}\ ,
\end{eqnarray}
where  $G_{(3)}=F_{3}-\tau H_{3}$ is the complex three-form (with $G_{(3)}=dC_{(2)},\ H_{(3)}=dB_{(2)}$),
$
\tau~=~C_{(0)}+ie^{-\phi}
$
is the complex IIB scalar and
\be
\tilde F_{(5)}=F_{(5)}-\frac{1}{2}C_{(2)}\wedge H_{(3)}+\frac{1}{2}B_{(2)}\wedge F_{(3)}\ ,
\ee
with the five-form $F_{(5)}=dC_{(4)}$. The equations of motion resulting from the
 above action have to be supplemented by the self-duality condition
\be
\tilde F_{(5)}=*\tilde F_{(5)}\ . \ee
In order to describe a background geometry sourced by $D3$-branes, we will consider the
following background warped metric
\ben ds^2 = h^{-1/2}ds_{\|}^2+ h^{1/2}ds^2_\bot\,,
\label{metricgeneric}
\een
where $ds_{\|}^2$ is the four dimensional metric along the
$D3$-branes and $ds^2_\bot$ the metric of the six dimensional
transverse space. In the case of vanishing three-form $G_{(3)}=0$,
the corresponding RR $4$-form gauge potential is given by
\ben
C_{(4)}=\left(1-\frac1{h}\right)dx^0\wedge\cdots\wedge
dx^3\,, \label{RR}
\een
while the dilaton field $\phi$ as well as the string coupling are
constant. The supergravity equations coming from the variation of
the action (\ref{action0}) are satisfied for Ricci flat $D3$'s if
and only if $h$ satisfies
\be h^{-1}\nabla_\bot^2h=0.\label{h} \ee
Moreover, supersymmetry requires that the four dimensional slice
parallel to the stack of $D3$-branes is flat Minkowski space-time,
whereas depending on the number of supersymmetries, the transverse
space can be flat $\mathbb{R}^6, ~\mathbb{R}^2 \times
\mathbb{H}_4$ ($\mathbb{H}_4$ hyperk\"ahler), Calabi-Yau (CY)  or a
generic Ricci-flat space $V_6$. In these cases, the $D3$-stack
breaks $1/2,1/4,1/8$ or all of the supersymmetries respectively,
leading to ${\cal{N}}=4,2,1$ or ${\cal{N}}=0$ four dimensional
theories.

We will introduce our $D3$-brane universe in the background
(\ref{metricgeneric},\ref{RR}) (for type-0 backgrounds see
\cite{Papantonopoulos:2004au}). In order to find analytical
results, we will rely on the probe-background approximation, in
which the backreaction of the probe brane onto the bulk fields is
disregarded. In order to ensure the validity of this
approximation, we need to keep the strength of the perturbation
produced by the probe small enough compared to the strength of the
source, as we shall discuss later. In particular, our $D3$  probe
universe is embedded along a four dimensional slice described by
the embedding fields $X^A(\xi^\mu)$, $A=0,\ldots, 9$, in term of
its local world-volume coordinates $\xi^\mu$, $\mu=0,\ldots, 3$.
Lengths along this slice are measured with the corresponding
induced metric, which is given in terms of the bulk metric and the
embedding fields by the pullback formula
\be
\label{induceed}
ds_{i}^2=g_{AB}\partial_\mu X^A \partial_\nu X^B d\xi^\mu d\xi^\nu\,.
\ee
Such a probe will then experience forces due to the background and
will consequently move through the bulk, with dynamics governed by
the Dirac-Born-Infeld action with a Wess-Zumino term
\ben
S_{DBI}+S_{WZ}= -T_3\int e^{-\phi} \sqrt{-g_i}\,d^4\xi - T_3\int C_{(4)} \, .
\label{actionn}
\een
The sign of the Wess-Zumino term has been chosen so as to represent a $D3$-brane, and $T_3$ is
the tension of the probe, given by
\be
T_3=\frac1{(2\pi)^3g_sl_s^4}\ .
\label{tension}
\ee
If there is additional
matter living in the brane, its contribution has to be added to
action (\ref{actionn}). However,
we will assume that the contribution to the probe motion of any additional matter
living on the brane is subdominant,
i.e. $T_3\gg \rho_m$, where ${\rho}_m$ is
the matter canonical energy density on the probe.

The induced metric (\ref{induceed}) will evolve during the probe
brane motion. From the point of view of an observer living in the
probe, such induced metric describes the evolution of his/her Universe.
We will introduce next an explicit description of a probe brane in a type
 IIB supergravity background , which will be
later used for our Slingshot scenario.

\section{Wandering $D3$-branes in warped throats}
\label{wandering}

{\noindent \bf {Background}}
$Dp$-branes may exists on compact manifolds \cite{Giddings:2001yu}. In the case
of CY compactifications, the CY manifolds may have singularities at
special points of their moduli space. Near a singularity, CY
looks like a conifold $\mathcal{M}_6$ on which, although singular, strings may
consistently propagate. The conifold geometry \cite{Candelas:1989js} is that of a
cone with a $\mathbb{T}^{1,1}$ base, whose topology  is
$S^3\times S^2$. At the singularity (the tip of the cone) both
$S^2$, $S^3$ shrinks to zero size. One may then introduce
a stack of $D3$-branes at the tip of the cone~\cite{Kachru:1998ys}, \cite{Kehagias:1998gn}, \cite{Klebanov:1998hh},
\cite{Acharya:1998db}, and the geometry  looks like  $AdS_5\times \mathbb{T}^{1,1}$ near the
$D3$-branes with a warp factor $h$.
Away from the $D3$'s,
the warped conifold is no longer a good description of the geometry. The
conifold singularity may now be deformed by blowing up the tip of the cone to an $S^3$ by
means of appropriate fluxes giving rise to the Klebanov-Strassler
geometry \cite{Klebanov:2000hb}. The latter provides  an IR description (close to the tip)
and it can be smoothly glued into a CY manifold in the UV (far from the tip)~\cite{Giddings:2001yu}. In
this case, we will expect that a wandering probe $D3$-brane will experience
mirage cosmology as it fall down towards the tip of the warped
deformed conifold in the Klebanov-Strassler throat \cite{Kachru:2002kx}.

In what follows we will need a detailed description of the motion of
our wandering probe $D3$-brane inside the above described throat,
assuming that at some point the brane leaves the throat reaching the
CY space. As to that end is not important the detailed geometry of
the conifold, we may replace it by flat six dimensional space,
(which corresponds actually to a $\mathbb{T}^6$
compactification~\cite{Verlinde:1999fy}, \cite{Chan:2000ms}), postponing the
more complicated case of the  deformed conifold to section \ref{ks}.
In this simplified configuration, the solution  is the maximally
supersymmetric one with flat Minkowskian four dimensional slice and
Euclidean transverse space with metrics
\be ds_\|^2=-d\eta^2+d\vec x \cdot d\vec x\, , ~~~ds_\bot^2 = d\vec
r\cdot d\vec r=dr^2+r^2\d\Omega_5^2\, , \label{metricgenericinside}
\ee
respectively, where $\d\Omega_5^2$ is the metric on an $S^5$. In
this case, the $h$ factor corresponds to the solution of (\ref{h})
sourced by a stack of $N$ $Dp$-branes sitting at $\vec r=0$. If the spacetime is chosen to be asymptotically Minkowski we have
\be
h=1+L^4/r^4\, ,\label{Min}
\ee
whereas if the spacetime is chosen to be $AdS_5\times S^5$
\be h=L^4/r^4\, .
\label{ads} \ee
We note that the two spacetimes are coincident in the deep throat region for which $L/r\ll 1$.
In both
cases, $L$ is related to
the RR charge $N$ of the $D3$-branes by
\be
L^4=4\pi l_s^4 N g_s\, .
\label{L}
\ee
Here $l_s$ is the string length, $g_s$ the string coupling and $N$ the number of
$D3$-branes in the stack.
The supergravity approximation we are using here are valid as
long as string perturbation theory can be applied and $\alpha'$
corrections are negligible, meaning the curvature radius $L$ of the
solution is big compared to the string length $l_s$, or equivalently
$g_s\ll 1$ and $g_sN\gg 1$.
\vspace{.1cm}

{\noindent \bf{Wandering} $D3$-brane} We now introduce a $D3$ probe with world-volume
coordinates $(\xi^0,\xi^i)$ in the vicinity of the $N$ coincident
$D3$-branes. In order to keep the validity of the probe-background approximation, we shall require $N\gg 1$.

In the following, we assume that the probe brane is extended parallel to the stack of
$D3$-branes so that it looks like a point moving in transverse space (for
inhomogeneous embedding see \cite{Galfard:2005rx}).
In this case, near the stack of
$D3$'s, the geometry is that of an $AdS_5\times S^5$ space. In the static gauge
$\eta=\xi^0, x^i=\xi^i$, with $\vec{r}=\vec{r}(\eta)$, the induced metric
is  given by
\be
ds_i^2=h^{-1/2}\left[-\left(1-h\vec{r}{\phantom{z\!}}'^2\right)d{\eta}^2+ d\vec x\cdot d\vec x\right]\ ,
\label{induced}
\ee
where a prime $(')$ is a derivative with respect to $\eta$.

Under the above described assumptions, the brane action
turns out to be
\bena
S &=& -\int {\cal L}d\eta\ , \label{action}
\eena
where the Lagrangian is given by
\bena
{\cal L}=-T_3V_3 \frac1h\sqrt{1-h {\vec{r}}{\phantom{z\!\!}}'^2}
- T_3V_3 \left(1-\frac 1h\right)\, ,
\eena
and $V_{3}$ is the un-warped volume of longitudinal brane directions (parallel to the  probe).

To begin the analysis of the motion,  we will use the rotational
symmetry of transverse space to write ${\vec{r}}'^2 = r'^2 +
r^2{\Omega'}_5^2$. Here ${\Omega'}_5$ represents
the angular velocity on the transverse $5$-sphere. From
the action (\ref{action}), we can define the conserved quantities
\be
\ell=\frac{\partial {\cal L}}{\partial \Omega_5'}=T_3 V_3\frac{r^2}{\sqrt{1-h {\vec{r}'^2}}}\Omega'_5
\ee
and
\be
{\cal E}=\frac{\partial {\cal L}}{\partial \Omega_5'}\Omega_5'+\frac{\partial {\cal L}}{\partial r'}r'-{\cal L}=
\frac{T_3 V_3}{h\sqrt{1-h {\vec{r}'^2}}}+T_3 V_3\left(1-\frac{1}{h}\right)\ .
\ee
These quantities parameterize the physical angular momentum and the physical brane energy, so we will loosely
call them angular momentum ($\ell$) and energy ($\cal E$).
Inverting the above equalities we obtain \cite{Kehagias:1999vr}
\be
\label{rd}
\Omega'_5
=
\frac{lL}{r^2h(1-C+U)}\ ,\ \ \
r'^2
= -V\, ,
\label{tres}
\ee
where
\be
\label{V}
C = 1-\frac1h\, ,
\ \ \ \ \ \ \ \
V=
-\frac 1h\left[1-\frac{1}{h^2(1-C+U)^2}\left(1+\frac{l^2 hL^2}{r^2}\right)\right]\, .
\ee
We have redefined the energy and the angular momentum as $U =
{\cal E}/T_3V_3-1$ and $l = \ell/T_3V_3L$, respectively.

In the
above formulae we note that the allowed regions for the motion
are those where $r'^2\geq 0$; the points where $r'^2=0$ being
the bouncing points.

We may  define the  cosmic time $t$ according to
\be
\frac{d\eta}{dt}=\frac{h^{1/4}}{\sqrt{1-h{\vec r}'^2}}=h^{5/4}{(q(C-1)+U)}\,  .
\label{proper}
\ee
The induced metric (\ref{induced}) can  now be written in the zero spatial curvature Friedman-Robertson-Walker form
\ben
ds_i^2=-dt^2+a^2(t)\,d\vec x\cdot d\vec x \,,
\een
where the scale factor is
\ben
a(t)=h^{-1/4}(r(\eta(t)))\, .
\label{scale}
\een
An observer living on the probe brane will therefore experience a cosmological
evolution parameterized by the two dimensionless parameters $l$ and $U$
that specify the form of the probe orbits. He/she may wonder
how much fine tuning is needed on these parameters in order to have
phenomenologically acceptable results.
\begin{figure}[t]
\centering
\includegraphics[angle=0,width=4in]{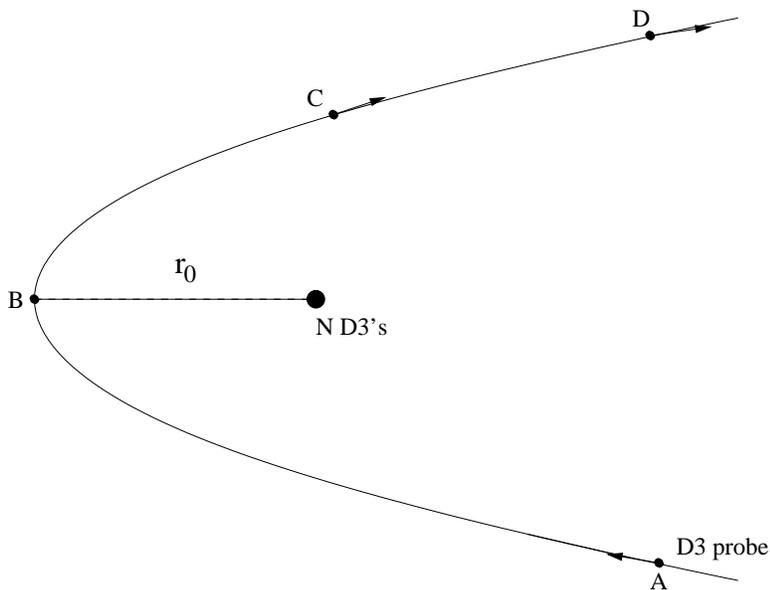}
\caption{Open orbit of a $D3$-brane probe in the background of N
$D3$-branes. The  probe starts at point A with some initial
velocity, reaches the turning point B and continues towards point D.
In the interval AB we have cosmological contraction on the probe and
expansion in the interval BD. Accelerated expansion exists for the
part BC of the orbit. \label{figu}}
\end{figure}
We are mainly interested here in the $l\neq0$ case, which
corresponds to a probe starting at $r=r_{in}$, with some initial velocity $r'_{in}$ (c.f. point A of
Fig.\ref{figu}) and traveling towards the stack of $D3$-branes at $r=0$.
During this travel, the velocity of the probe decreases until the turning
point B at $r=r_b$, where the velocity of the probe vanishes. After
that point, the outgoing probe velocity increases again until it
reaches its initial value at point D.

As we are now going to show, in the case where the probe brane
follows an orbit around the stack with a turning point, an observer
on the brane will measure a cosmological accelerated expansion
during a short period in the outgoing part of the orbit (interval BC of Fig.\ref{figu}) and a
decelerated expansion from that point on. With the above
definitions, the expansion rate is
\ben
\dot a = \frac {da}{dr}\, r'
\,\frac{d\eta}{dt}\ ,
\label{uno}
\een
where we denote with a dot $(\dot{})$ derivatives with respect to
cosmic time $t$.

An observer in the brane will measure
cosmological contraction during the ingoing part of the orbit,
an expansion during the outgoing part, and a bounce at the turning
point $r_b=r(\eta_b)$ such that $r'(\eta_b)=0$. We can readily see that the cosmic
acceleration can be expanded around the turning point as
\be
\label{acc}
\ddot a
\propto r' \,\frac{dr'}{dr}+ {\cal O}(r'^2)\, .
\ee
When the probe brane
approaches the bouncing point $r_b$, $r$ decreases while the
brane is climbing a centrifugal barrier so that $r'$
decreases as well. Conversely, when $r$ is growing, $r'$ is also growing,
as the brane is falling down the barrier. This  implies that
$dr'/dr$ is positive and therefore  $\ddot a$ is negative in the vicinity
of the turning point $r_b$ in the incoming branch of the orbit, and
it becomes positive in the outgoing branch. Consequently, the brane
observer will measure cosmological acceleration when the brane has
passed through the turning point as it moves away from the stack of
$D3$'s. It is therefore interesting to ask how many e-folds such an
observer will experience during the accelerated expanding era. The
number of e-folds is defined as $ N_e=\ln(a(\eta_{f})/a(\eta_{b}))$
where $\eta_{f}$ is the time in which the acceleration ends, and
$\eta_{b}$ the inversion time. In order to have as many e-folds as
possible it would be therefore important to have
$a(\eta_{f})=h^{-1/4}(r(\eta_f))$ as big as possible, or
equivalently, $r(\eta_{f})$ as big as possible together with the
requirement that $r(\eta_{b})$ is as small as possible.

To take this analysis further, we need some explicit knowledge on the
orbits and the resulting cosmological evolution.
\subsection{Asymptotically Minkowski Throat}
\label{flatt}

It can be easily seen that in the  case that the warping factor $h$
in the metric (\ref{metricgeneric}) is chosen by Eq.(\ref{Min}),
flat Minkowskian spacetime is recovered in the large $r\gg L$ limit.
In this background, the probe brane  will feel some effective
potential when it is close enough to the stack of $D3$'s, while it will move more or less freely when it is far from
it. We will study the motion of the probe and the resulting
cosmology in this section.

We are interested in unbounded orbits, which exists as long as
$V<0$ at infinity. Solving Eq.(\ref{tres}) for $r'^2$ we get that
the allowed region for the motion ($r'^2 \geq 0$) is
\be
\label{ri}
r^2\geq\frac{L^2}{2U(U+2)}\left[l^2+\sqrt{l^4-4 (U+2)U^3}\right]\ .
\ee
In order to have a bounce, angular momentum and energy are therefore constrained to be
\be
\label{zero} l^4-4 (U+2)U^3\geq 0\ .
\ee
In this case an inversion point $r_b$ such that $r'=0$ always exist
at the value of $r$ that saturates the inequality (\ref{ri}).
Then the probe follows an unbounded orbit coming from infinity,
bouncing at $r_b$, and then going back to infinity. The explicit
form of these orbits can be obtained by quadratures integrating
$d\Omega_5/dr=\Omega_5'/r'$ \cite{Burgess:2003qv}.

In order to maximize $N_e$, the acceleration
period should end when the probe is far away from the stack of $D3$-branes, or in other words
in the $r_f\gg L$ region. We can already anticipate that this cannot
work as in the asymptotic region, close to the hat of the CY, the probe brane does not feel any potential.
A numerical exploration of the $(l,U)$
plane, indeed shows that the $N_e$ is not bigger than $\sim 0.4$.
Also analytical arguments can be given as follow.

The condition to have a positive acceleration in the $r_f\gg L$ region can be found by expanding the acceleration in the $\epsilon=r_f/L$
parameter, so to have
\be\label{contr}
\frac{l^2L^2}{r_f^2}>\frac{10}{7}(U+2)U\ .
\ee
However, from the $r'^2>0$ constraint we get
\be
(U+2)U>\frac{l^2L^2}{r_f^2}\ ,
\ee
which is obviously in contradiction with (\ref{contr}). So, the acceleration cannot finish in the flat region as anticipated.

We conclude that the whole acceleration process must take place in
the throat region. We will now be more explicit in the
next paragraph, studying the motion of the probe inside the
$AdS_5\times S^5$ throat.

\subsection{$AdS_5\times S^5$ space}
\label{adss}
In this background the warping factor $h$ is given by
Eq.(\ref{ads}) and therefore $a=r/L$. Again we can use Eq.(\ref{tres}) to understand  the
orbits of the probe brane. In order to have unbounded orbits ({\em
i.e.} in order to let our probe to eventually leave the throat
instead of being confined in it), we again impose that $V$ is non positive at
infinity. This leads us to the choice $U>0$. From the condition $r'^2\geq 0$
we get that the allowed region for the brane motion is
\be
\label{riAdS} r^2\geq\frac{L^2}{4U}\left[l^2+\sqrt{l^4-8
U^3}\right]\ , \ee
and for a bouncing point we should have
\be
l^4-8 U^3>0\, .
\label{laspe}
\ee
In this case, a turning point $r'=0$ always exists at
$r=r_b$ where the inequality (\ref{riAdS}) is saturated.
Thus, when the condition (\ref{laspe}) is fulfilled, the probe brane
coming from infinity bounces at $r_b$ and then goes back to
infinity.

Requiring $\ddot a >0$ and $r'^2>0$ we
obtain an upper bound for the accelerating region
\be
\label{rlim}
r^2\leq\frac{L^2}{2 U}\left[{l^2+\sqrt{l^4-6 U^3}}\right]\ \  \mbox{ with}\ \ddot a>0\, .
\ee
It is then simple to see that the number of e-folds is
\be
N_e= \frac12\log\left(2\frac{{l^2+\sqrt{l^4-6 U^3}}}{{l^2+\sqrt{l^4-8 U^3}}}\right) \lesssim 0.7\ ,
\ee
consistently with the numerical results cited above.

Then the cosmological evolution as seen for an observer in the brane
does have an inflationary period, but it is too short to be used for
the resolution of the standard cosmological problems. Nevertheless, we will show in section \ref{realistic} that horizon, isotropy and curvature
problems can be naturally solved in our proposed Slingshot scenario.

\subsection{Non-vanishing spatial curvature}
\label{curved}

The flatness of the Universe is one of the vexing problems of standard cosmological scenarios.
In the previous discussions we considered embedding with spatially flat three-sections. However, Friedman geometries with non vanishing three-curvature can also be embedded on a maximally symmetric bulk.
To this
end, we will consider again an $AdS_5\times S^5$ background. An appropriate change of coordinates puts the metric (\ref{metricgeneric}) into
the form
\be
\!\!\!\!\!\!\!\!\!\!\!\!\!\!\!\! ds^2=-(\kappa+\frac{r^2}{L^2})d\eta^2+\frac{dr^2}{\kappa+\frac{r^2}{L^2}}+
\frac{r^2}{L^2(1+\frac{\kappa}{4L^2}\rho^2)^2}\left(d\rho^2+\rho^2d\Omega^2_2\right)
+L^2d\Omega^2_5\ .
\ee
The constant $\kappa$ parameterizes the flat, spherical or hyperbolic three-curvature of the four dimensional slice, and can be set respectively to $\kappa=0,1,-1$ by the following re-scaling of coordinates: $\rho\rightarrow {\rho}/{\sqrt{|\kappa|}}\ ,
r\rightarrow{r}{\sqrt{|\kappa|}}\ , \eta\rightarrow {\eta}/{\sqrt{|\kappa|}}$.

The RR $5$-form field strength is
\be
F_{(5)}=-\frac{4}{L}(1+*)Vol(AdS_5)\ .
\ee
The corresponding potential is given, after  a straightforward integration in the $r$ variable, by
\be
C_{(4)} = \left(1-\frac{r^4}{L^4}\right)\frac{\rho^2}{(1+\frac{\kappa} {4L^2}\rho^2)^3} d\eta\wedge d\rho\wedge d\Omega_2\ ,
\ee
where, without loss of generality, we have set the integration constant to 1.

The probe with
time-like coordinate $\xi^0$ and spatial polar coordinates
$(\xi,\Xi_2)$ can be embedded along the new coordinates
$\xi^0=\eta,\xi=\rho$ and $\Xi_2=\Omega_2$, at a time-dependent position
$r(\eta),\Omega_5(\eta)$ in transverse space.
The induced metric on the brane is
\be
\!\!\!\!\!\!\!\!\!\!\!\!\!\!\!\!\! ds_i^2=-\left[\kappa+\frac{r^2}{L^2}-\frac{r'^2}{\kappa+\frac{r^2}{L^2}}-L^2\Omega_5'^2\right]d{\eta}^2+
\frac{r^2}{L^2(1+\frac{\kappa}{4L^2}\rho^2)^2}\left(d\rho^2+\rho^2d\Omega^{ 2}_2\right)\ ,
\label{indcurv}
\ee
and therefore the DBI action turns out to be
\be
\!\!\!\!\!\!\!\!\!\!\!\!\! S = -T_3V_3\int d\eta \frac{r^3}{L^3}
\sqrt{\kappa+\frac{r^2}{L^2}-\frac{r'^2}{\kappa+\frac{r^2}{L^2}}-L^2\Omega_5'^2}
-T_3V_3\int d\eta \left(1-\frac{r^4}{L^4}\right)\, .
\ee
Following the same procedure as in the previous paragraphs we find
\be
\Omega_5'=\frac{l}{L(1-C+U)}\left(\kappa+\frac{r^2}{L^2}\right)\ , \ \ \ \ \ r'^2=-V\ ,
\ee
and
\be
V=-\left(\kappa+\frac{r^2}{L^2}\right)^2\left[1-
\frac1{(1-C+U)^2}\left(\kappa+\frac{r^2}{L^2}\right)\left({l^2}+\frac{r^6}{L^6}\right)\right]\ .
\ee

With this in hand, we can now investigate the form of the orbits for non-vanishing $\kappa$. It is not difficult to verify that
\be
r'^2 \propto  U^2 - \kappa l^2 -l^2 x -2U x^2 -
\kappa x^3\ , \label{rio}
\ee
where $x=r^2/L^2$ and the proportionality function is strictly positive. The zeros of the cubic polynomial (\ref{rio})
in the positive $x$ semi-axis correspond to the bouncing points for our probe brane motion.
For negative $\kappa$, in the  $x\to\infty$ limit of Eq.(\ref{rio}), we have
 $r'^2\sim -\kappa x^3>0$, the polynomial is positive
there and infinity is part of the allowed region. Conversely, at $x=0$ we have
$r'^2\sim U^2-\kappa l^2>0$ and the origin is also part of the allowed region.
Since the slope of the polynomial is negative at $x=0$, there exist either two
or zero positive roots, according to the sign of the cubic discriminant. When the
discriminant is positive, we will necessarily have two positive roots, and
then two bouncing points, while when it is negative, the polynomial reaches
a minimum positive value and then grows to infinity with no positive roots.
We conclude that for negative discriminant, the probe brane can move from infinity to the origin
without finding any bouncing point ({\em i.e.} hitting eventually
the stack). On the other hand,  for positive discriminant there are ``inner''
orbits hitting the stack with an outer bouncing point and
``outer'' orbits coming from infinity, bouncing at some fixed radius
$r_b$ and then going back to infinity.

For positive $\kappa$, we find $r'^2\sim -\kappa x^3<0$ at infinity, which
implies that infinity is not part of the allowed region.  Thus,  all
orbits are bounded, {\em i.e.} there always exists an outer bouncing
point. On the other hand, for $4U^2-3\kappa l^2>0$ there
are two zero slope points in the positive $x$ axis, meaning
that the curve goes through the vertical axes with negative slope.
Then, when the discriminant is positive two of the roots should be at
the positive $x$ axis so that  there is also an inner bouncing
point. In this last case, all orbits are bounded.

Even if the explicit expression of the cubic discriminant is not
very enlighten, it can be checked that it is positive for large
enough angular momentum, and we will assume this in what follows.
Whenever we need an explicit lower bound for $l$, we will use the
expression (\ref{laspe}) for the case of zero curvature as an
estimation of it.

The conclusion is that for large enough angular momentum, when $\kappa<0$ the
probe moves coming from infinity up to some finite distance to the
origin $r_b$ and then goes back to infinity. On the other hand, when $\kappa>0$
there is an additional outer bouncing point preventing the
probe to go too far from the origin. This result can be translated
into the familiar statement that positive curvature universes  will
expand up to some finite scale parameter and then contract, while
negative curvature ones expand forever.

What kind of cosmological evolution would we get with this modified
scenario? To answer that we change variables in the above induced
metric (\ref{indcurv}) to get
\be
ds^2_i=-dt^2+\frac{a(t)^2}{(1+\frac \kappa{4L^2}\rho^2)^2}(d\rho^2+\rho^2d\Omega^2_2)\ ,
\ee
where again the scale factor is given by $a=r/L$ and the cosmic time has been defined according to
\be
\frac{d\eta}{dt}=\frac{L^3(1-C+U)}{r^3\left(\kappa+\frac{r^2}{L^2}\right)} \ .
\ee
Note that now our cosmological model is specified giving two
continuous $l,U$ and one discrete $\kappa$ dimensionless parameters.

Although we again have an accelerating period at the turning point of the brane orbit, the same numerical analysis as
before shows that the number of e-folds is again ridiculously small.

\section{Slingshot cosmology: a realistic scenario}
\label{realistic} In the previous sections we described the cosmic
evolution as experienced by a brane observer while the $D3$-brane is
probing the bottom of the CY throat. We consider this evolution as
describing the early time cosmology of our Universe. We will
consequently assume that at some point, the orbit reaches the top of
the throat and moves into the CY space to which the throat is glued.
As the brane moves away from the throat, mirage approximation breaks
down and the  dynamics is determined by local gravity on the brane.
This is the analog of the exit of the inflationary era in standard
inflation. Unfortunately, the explicit description for the
transition from the mirage to local gravity dynamics is lacking at
the moment. Intuitively what happens is that, at the bottom of the
throat, very close to the heavy stack of $N$ branes sourcing it, a
slowly moving additional brane has a negligible effect on the
background as compared with the distortion produced by the stack,
for $N\gg 1$. On the other hand, when the brane is far from the
stack and it leaves the throat, there is no other heavy object
around, and the backreaction of the brane and of any matter living
on it, becomes the only important correction to the background
metric. At this point backreaction causes expansion of the 4d slice,
and in consequence the brane motion is Hubble damped. Four
dimensional local gravity might then be realized as in
\cite{Shiromizu:1999wj},\cite{CottaRamusino:2006iu}.

Inside the throat, the motion is completely described by the energy $U$ and
angular momentum $l$ of the orbit,  the number $N$ of branes in the stack  and the sign of $\kappa$.
The natural question is whether we can  choose these free parameters of our model so as to have a realistic cosmology in the probe brane and, in case we can, how much fine tuning is needed on these parameters.

As we have found in the previous sections, our cosmological model passes through an inflationary era, but it is very short and it
only provides a small amount of inflation ${\rm N}\leq 0.7$. Is this reason enough to rule it out? To answer that, we must reexamine
the naturalness problems that originally inspire inflationary scenarios, and check if they are also present in our Slingshot model.

\subsection{Early Time Cosmology}
\label{early}

To gain intuition in the behavior of the scale factor, it is now convenient to calculate the Hubble constant $H^2=\dot a^2/a^2$ and write Hubble
equation using (\ref{tres}) and (\ref{proper})
\be
\label{h2}
H^2=-\frac1{L^2}\left[\frac{\kappa}{a^2}-\frac{2U}{a^4}+\frac{l^2}{a^6}+\frac{\kappa l^2-U^2}{a^8}\right]\ .
\ee
Here we see that mirage matter behaves like a curvature term, a
radiation term (this is the ``dark radiation" of \cite{Binetruy:1999hy}, or the mirage radiation of \cite{Kehagias:1999vr})
 and some higher order terms depending on energy and
angular momentum. At late times the higher order terms are subdominant, ensuring that
we can smoothly match the evolution with a standard cosmology in the local gravity era.

\vspace{.1cm}
\noindent{\bf Homogeneity and Isotropy problem} In the previous sections we have established that for large enough
angular momentum the orbits described by the probe have a bouncing
point. When this condition is satisfied, the probe brane never reaches
the origin $r=0$ where the scale factor $a=h^{-1/4}$ vanishes. Then
we realize that in the brane Slingshot scenario there is no
singularity and consequently there is no horizon problem. A
different way to say that is to note that since there is no big
bang, the universe has had time enough to reach any desired degree
of homogeneity. To be more precise, standard arguments \cite{KolbTurner} require the comoving
horizon to be bigger than the Hubble horizon today $H^{-1}_0$ in order to solve the horizon problem. In fact in this case
each point on the last scattering surface was causally connected sometime in the past.

The comoving horizon is defined as
\be
\Delta\eta=\int^{\eta_0}_{\eta_{i}}d\eta'\ ,
\ee
where $\eta_0$ is the conformal time today and $\eta_i$ is the smallest conformal time in the
Universe evolution. As we have a bounce, $\eta_i\rightarrow -\infty$ and the condition to
solve the horizon problem
\be
\Delta\eta\gg H^{-1}_0\ , \label{comoving}
\ee
is trivially satisfied.

Another important problem in standard cosmology is the isotropy problem, which
can be  formulated as follow: A measure of anisotropy is the shear. The shear is defined as
$\sigma_{\mu\nu}=h^\alpha_\mu h^\beta_\nu \nabla_{(\beta} u_{\alpha)}$, where $u_\alpha$ is the four
velocity of a timelike geodesic observer and $h_{\alpha\beta}=g_{\alpha\beta}+u_\alpha u_\beta$ is the
three metric orthogonal to the four velocity $u_\alpha$. In a FRW background
$\sigma^2=\sigma_{\alpha\beta}\sigma^{\alpha\beta}\sim a^{-6}$. Purely geometrical considerations
imply that for an anisotropic perturbation we have \cite{Barrow:2004he}
\be
H^2=\sigma^2+\frac{8\pi G_N}3\rho\ ,\label{shear}
\ee
where $\rho$ is the matter energy density, and it satisfies energy conditions, $\rho\sim a^{-n}$ where $n\leq 4$ and $G_N$ is the gravitational coupling to brane matter.
In this case, when the scale factor is close to the singularity, shear dominates making the universe more and more anisotropic.
This kind of behavior also generate a chaotic evolution at early time making the Universe very unstable under small
perturbations. In other words, in order to produce the small anisotropy we observe today, standard cosmology needs an extreme fine tuning.

In the Slingshot this problem is circumvented as in Cyclic Scenarios \cite{\cite{Wesley:2005bd}}. As we shall see later, the non-relativistic
limit for our Universe evolution, does not allow the scale factor to reach very small values. In this way the shear cannot
dominate. But even if this condition is relaxed, the angular momentum term $l^2a^{-6}$ will always dominate over the shear
perturbation. Moreover, if angular momentum is negligible, as mirage matter violates energy conditions, the Hubble
equation (\ref{h2}) contains terms scaling like $a^{-8}$. These terms obviously dominates with respect to the shear
at high energies (small $a$ values) avoiding the Kasner behavior typical of standard cosmology close to the singularity.
Therefore the isotropic solution we used is stable under anisotropic perturbations.

\vspace{.1cm}
\noindent{\bf Flatness Problem}
In our model, the spatial curvature is a function of our orbital parameters
$U,l,\kappa$ that specify the orbit and the embedding of the probe
brane. However, it is important to see whether these parameter have
to be fine tuned or not in order to obtain phenomenologically
acceptable results.

To that end, we will find the minimum of the spatial curvature using (\ref{h2}).
We will assume $U\gg l$ in order to disregard the $\kappa l^2/a^8$ term. We will check that this
assumption is actually satisfied at the end of the calculation. Furthermore we will work in
the $U/a^4\ll 1$ limit, and then discard the $U^2/a^8$ term. As we will explain later, this
corresponds to non-relativistic motion of the probe brane. With these assumptions,  (\ref{h2}) becomes
\be
H^2=-\frac1{L^2}\left[\frac{\kappa}{a^2}-\frac{2U}{a^4}+\frac{l^2}{a^6}\right]\ .
\ee
The curvature term can be discarded at late times if it passes through a minimum at a very small value of the quantity
\be
|\Omega-1|=\frac1{L^2a^2\bar H^2}\ ,\label{omegaminusone}
\ee
where
\be
\bar H^2=-\frac1{L^2}\left[-\frac{2U}{a^4}+\frac{l^2}{a^6}\right]\ ..\label{37}
\ee
It can be easily verified that the minimum of (\ref{omegaminusone}) is at $a^2 = {3l^2}/{4U}$, where it takes the value
\be
\frac 1 {L^2a^2\bar H^2} \simeq \big{(}\frac lU\big{)}^2\ll 10^{-8}\ ..\label{ornitorrinco}
\ee
The last inequality has been imposed in order to get an agreement with observation. This is satisfied whenever
\be
l\ll  10^{-4}U\ ,
\label{ineq}
\ee
consistently with our approximations.

We now combine this limit with the limit (\ref{laspe}) in order to have the bouncing, we obtain the necessary condition
\be
\label{E}
U\gg 8\times 10^{16}\ .
\ee
The conclusion is that observational constraints in the curvature are enforced in our model by (\ref{ineq}).
Note that, as can be seen in Fig.\ref{parameters}, even satisfying
all the constraints (\ref{ineq}) and (\ref{E}) we still have an
enormous region of parameter space in which to make our choice, and in this sense the fine tuning problem is relaxed.

\subsection{Density Perturbations}
\label{perturbations}
In order to compute density perturbations in the present setup, we will use the non-relativistic approximation. This is equivalent to  adiabatic mirage cosmological expansion and implies that $h
(r'^2+r^2\Omega'^2)\ll 1$. From Eq.(\ref{rd}) we get that
\be
\label{non}
h(r'^2+r^2\Omega_5'^2) =
 1-\frac{1}{h^2\left(1-C+U\right)^2}\ .
\ee
Therefore a sufficient condition to satisfy the non-relativistic approximation is
\be
1-\frac{1}{h^2\left(1-C+U\right)^2}\ll 1\, \label{nr},
\ee
or
\be
2 h U\ll 1\ . \label{nonrel}
\ee
This condition could have been already guessed before. In fact, the physical brane energy is $hU$,
therefore the non-relativistic approximation consistently
requires (\ref{nonrel}).
Note that Eq. (\ref{nr}) is equivalent to the condition $U/a^4\ll1$
that we used in the previous section to study the flatness
problem.

With this approximation the calculations can be made explicit
and we will be able to find the power spectrum for the cosmological
perturbations. We will suppose as usual that the gravitational perturbations,
in our case produced by quantum fluctuations of the brane embedding, will be
straightforwardly passed to the matter fields.

\vspace{.1cm}
{\noindent \bf{Power Spectrum}}
We are interested in the spectrum of the scalar fluctuations of
the gravitational field around its classical value. A straightforward
way is to allow the embedding fields $r,\Omega_5$ to depend on all
brane coordinates {\em i.e.} $r=r(\eta,x^i)$, $\Omega_5=\Omega_5(\eta,x^i)$. Then the induced metric turns
out to be
\begin{eqnarray}
\!\!\!\!\!\!\!\!\!\!\!\!\!\!\!\! \!\!\!\!\!\!\!\!\!\!\!\!\!\!\!\! ds_i^2=-\left(\frac{r^2}{L^2}-\frac{L^2r'^2}{r^2}-L^2
\Omega_5'^2\right)d\eta^2&+&\left(\frac{r^2}{L^2}\delta_{ij}+ \frac{L^2
\partial_i r\partial_j r}{r^2}+L^2\partial_i {\Omega_5}\partial_j {\Omega_5}\right)dx^idx^j+\cr
&+&2\left(\frac{L^2r' \partial_i r}{r^2}+L^2{\Omega_5}'\partial_i{\Omega_5}\right)d\eta
dx^i\ .
\end{eqnarray}
According to standard results \cite{Boehm:2002kf} the Bardeen potentials in the non-relativistic approximation are
\be
\Phi= -\frac{\delta r}{r}\, , ~~~~ \Psi=-\Phi\ .
\ee

It is generally assumed that the present power spectrum of scalar
fluctuations in the CMB as measured at WMAP has been produced by quantum
fluctuations in the early universe. The usual way to calculate this
spectrum is to quantize the appropriate fields and canonically
normalize them. After that, the quantum amplitudes are taken as an
initial conditions for a classical evolution up to present. The details
of the quantum to classical transition are subtle, and its explanation would necessarily
include some kind of decoherence mechanism or  ``collapse of the wave function'' as analyzed
in \cite{Perez:2005gh}. We will simply assume that such mechanism exists and that it is in
agreement with the above described picture.

We will first describe the classical dynamics of the perturbation on
the probe $D3$, expanding the DBI action up to quadratic order in
derivatives. In the
non-relativistic limit, which is equivalent to adiabatic expansion, the resulting DBI action is
\bena \label{actionharm} S&=&-
\frac{T_3}{2}\int d^4x
\left(
\partial_\mu r\partial^\mu r + r^2\partial_\mu\Omega_5\partial^\mu\Omega_5
\right)\ ,
\label{dbi}
\eena
where the metric and the integral measure are the flat
Mink\-owsk\-ian ones.

The equations of motion, which follows from the action (\ref{dbi}) are
\be
\partial^\mu\partial_\mu r=r\partial_\mu \Omega_5\partial^\mu \Omega_5\, ,
~~~~\partial^\mu(r^2\partial_\mu\Omega_5)=0\ .
\label{2}
\ee
For the unperturbed motion of the probe we have
$r=r(\eta),\Omega_5=\Omega_5(\eta)$ so that
\be
r''-r\Omega_5'^2=0\, , ~~~(r^2\Omega_5')'=0\ , \label{3}
\ee
and the conserved first integrals of this system are
\be
r'^2+\frac{l^2L^2}{r^2}= 2 U \, , ~~~
r^2\Omega_5'=lL\ ,
\label{enn}
\ee
where we have related them with the angular momentum and energy of the corres\-ponding relativistic problem,
 by expanding (\ref{tres}) up to quadratic order in time derivatives.
The solution to  Eq.(\ref{3}) is then
\be
r=\sqrt{2U\,\eta^2+\frac{l^2L^2}{2 U}}\, ,
~~~~~\Omega_5=\arctan\left(\frac{2U}{lL}\,\eta\right)\, ,\label{8}
\ee
where a constant of integration has been fixed by requiring that
at $\eta=0$ the probe is at the turning point
\be
r_b=\frac{lL}{\sqrt{2U}}\, . \label{rb}
\ee
One can easily find that the  inequality for the validity of the non-relativis\-tic app\-roximation  (\ref{nonrel}) is always fulfilled (with
$h=L^4/r^4$) for $2 U L^4/r_b^4\ll 1$, or
\be
{l^4}\gg{8 U^3} \ .
\label{nonrel1}
\ee
This is an important constraint and it should be satisfied in order our solution to be valid. It simply stretches  the fact that the brane
may keep always its slow velocity as it approaches and turns around the stack of the background D-branes.
Moreover, since (\ref{nonrel1}) is stronger than the bouncing
condition (\ref{laspe}), it ensures that in the non-relativistic approximation we
always have a bounce. This can be explicitly seen in the solution above. The induced
metric on the probe in the non-relativistic limit
turns out then to be
\be
ds^2=\left(\frac{2U}{L^2}\,\eta^2+\frac{l^2}{2U}\right)\Big{(}\!\!-d\eta^2+d{\vec{x}}{\,}^2\Big{)}\, ,  \label{dse}
\ee
where $\eta$ is the conformal time and clearly represents a bouncing universe.

At very early and late times ($|\eta|\gg l L/2U$) we have a radiation dominated
universe, while the initial singularity is avoided as the universe has minimal ``radius" $r_b$. This is also compatible with the fact that the
geometry (\ref{dse}) violates energy conditions. Indeed, the effective energy momentum tensor for (\ref{dse}) has non-zero components
\be
T_{\eta\eta}=\frac{48U^4 \eta^2}{(l^2 L^2+4 U^2 \eta^2)^2}\, , ~ ~ ~ ~
T_{ij}= - \delta_{ij}\frac{8(L^2 l^2-2U^2\eta^2) }{(l^2 L^2+4 U^2 \eta^2)^2}
\ee
and it can easily be seen that both, weak and strong energy conditions  are violated for
\be
| \eta|<\frac{Ll}{2\sqrt{2} U}.
\ee
We note that (\ref{actionharm}) is the action of a free two-dimensional harmonic oscillator. In fact, by using the change of variables
$X=T_3^{1/2}\, r\cos\Omega_5$ and $Y=T_3^{1/2}\, r\sin\Omega_5$ we have
\be
S=-\frac{1}{2}\int d^4x \left(\partial_\mu X\partial^\mu X+ \partial_\mu Y\partial^\mu Y\right)\ .
\ee
Then, the quadratic action for the perturbations $X+\delta X,\, Y+\delta Y$ is
\be
S_\delta=-\frac{1}{2}\int d^4x \left(\partial_\mu \delta X\partial^\mu \delta X+ \partial_\mu \delta Y\partial^\mu \delta Y\right)\ ,
\ee
or, in Fourier space
\be
S_\delta=\frac{1}{2}\sum_k\int d\eta \left(\delta X_k'^2+k^2\delta X_k+\delta Y_k'^2+k^2\delta Y_k\right)\ ,
\ee
where $k$ is the wave number of the fluctuation. These fluctuations are related with those in the original variables by
\begin{eqnarray}
T_3^{1/2}\, \delta r_k=\delta X_k\cos\Omega_5+\delta Y_k\sin\Omega_5\ ,\cr
T_3^{1/2} \, r\delta \Omega_k=\delta Y_k\cos\Omega_5-\delta X_k\sin\Omega_5\ .
\end{eqnarray}
The equation of motion for the variable $\delta X_k$ and $\delta Y_k$ are
\be
\delta X_k''+k^2\delta X_k=0\ , \ \delta Y_k''+k^2\delta Y_k=0\ ,
\ee
and the solution is the standard plane waves
\be
\delta X_k=A_x \sin(k\eta+\phi_x)\ ,\ \delta Y_k=A_y \sin(k\eta+\phi_y)\ .
\ee
Here, $A_x,A_y$ are constant amplitudes and $\phi_x,\phi_y$ initial phases.

It is now straightforward to verify that for $k\ll lL/r^2$, the Bardeen variable $\Phi=-\delta r_k/r$ is frozen, i.e.
$(\delta r_k/r)'\simeq 0$ in the limit $r\gg r_b$.
The same result could also be deduced in the original $(r,\Omega_5)$ variables as it is illustrated in the Appendix.

In order to calculate the power spectrum of perturbations we need
now some consideration on the quantum behavior of the brane
fluctuations. A given mode will behave quantum mechanically as long
as there is no ``collapse of the wave function'' occurring. The key
assumption usually made is that, at some point of the cosmological
evolution, the wave-function collapses and then the mode begins to
behave classically. We will keep this assumption here, but we will
choose the time at which this collapse happens differently: instead
of the usual horizon crossing, we will suppose that it happens when
the proper wavelength of the mode becomes larger than some
collapsing length $l_{c}$. This is the mechanism proposed in
\cite{Hollands:2002yb} to produce a flat power spectrum without inflation.
The need for such a collapse mechanism in any quantum theory of gravity was emphasized in \cite{Perez:2005gh}. It was also found in the discussion of transplankian effects in \cite{Bozza:2003pr}, and in the very different context of noncommutative inflation in \cite{Alexander:2001dr}.
Then we will say that classical modes are created at the time $\eta_*$
when the proper wavelength of the corresponding quantum mode reaches the value $a(\eta_*)/k\equiv a_*/k = l_c$.

In the quantum mechanical regime we have
\be
\delta X_k=v a_x+v^*a_x^\dag\, , ~~~~\delta Y_k=u a_y+u^*a_y^\dag
\ee
where
\be
u=v=\frac{e^{-ik \eta}}{\sqrt{2 k}}\ .
\ee
and therefore we have
\be
\frac{\langle\delta r_k r_{k'}\rangle}{r^2}=\frac{\delta_{k,k'}}{2T_3kr^2}\ .
\ee

The above quantum mean value for $\delta r^2_k/r^2$ will then evolve like $1 /{r(\eta)^2}$ up to the time at which the referred collapse takes place.
After that moment, it will behave classically and, according to the previous considerations, co-moving wave numbers smaller than the $lL/r^2$ curve
will be frozen. With that, for frozen modes, we have
\be
P=\frac{1}{2\pi^2}
\frac{|\delta r_k|^2}{r^2}=\frac{1}{2\pi^2}
\frac{|\delta {r}_k(\eta_*)|^2}{r_*^2}\ .
\label{ppp}
\ee
Then, we get an exact scale invariant spectrum
 \be
 P=\frac{2\pi g_s l_s^4}{L^2 l_c^2}  \frac{1}{k^3}\ ,\label{flat}
 \ee
due to the fact that
 \be
|\delta r_k (\eta_*)|^2=\frac{1}{2T_3k}\, , ~~~~r_*= L \,a_*=L \,k l_c\ .
\ee
In this way we have gone around the general arguments of \cite{Creminelli:2004jg} about the blue spectrum of bouncing cosmologies\footnote{See \cite{Nayeri:2005ck} for other examples of scale invariant spectrum in bouncing cosmologies.}.

\section{A Slingshot with $n_s\simeq .95$}
\label{KSsection}
In the previous section, we found an exactly flat ($n_s=1$) power spectrum for the perturbations. Nevertheless,
accordingly to WMAP measurements, the observed spectral index is slightly red $n_s\simeq .95$. Moreover, the
warped conifold geometry was replaced by the simpler $AdS_5\times S^5$ metric, which is not realistic from the point of
view of string compactifications.
In this section, we will show that both problems are closely related and in fact, using
the resolved conifold Keblanov-Strassler (KS) solution for the throat metric, the resulting
spectrum has a slightly red spectral index.
\label{ks}

\subsection{Orbits in the Klebanov-Strassler throat}
Let us consider here a $D3$-brane probe moving in the background of a warped throat region in a
CY compactification of type IIB string theory. Instead of approximating the throat with the conifold
geometry with a large number of $D3$-branes on its tip, we will resolve the singularity using
the Klebanov-Strassler (KS) warped deformed conifold, where
the tip of the conifold has been blown-up.
In this case the metric (\ref{metricgeneric}) has a transverse part given by the KS \cite{Kachru:2002kx} geometry
\be
\!\!\!\!\!\!\!\!\!\!\!\!\!\!\!\!\!\!\!\!\!\!\!\!\!\!\!\! d s_\perp^2\!\!=\!\!
\frac{\epsilon^{4/3}K}{2}\!\!\left(\!\frac{1}{3
K^3}\!\!\left[d\tau^2\!+\!(g^5)^2\right]\!+\!
\cosh^2\!\!\left(\!\frac{\tau}{2}\!\right)\!\!
\left[\!(g^3)^2\!+\!(g^4)^2\right]\!+\!\sinh^2\!\!\left(\!\frac{\tau}{2}\!\right)\!\!
\left[(g^1)^2\!+\!(g^2)^2\right]\!\right)\ .
\ee
Here $\epsilon$ is the resolution parameter, resolving the tip of
the cone, and $K$ is a function of the ``radial'' variable $\tau$
given by \be
K(\tau)=\frac{\left(\sinh(2\tau)-2\tau\right)^{1/3}}{2^{1/3}\sinh
\tau}\, . ~~~ \ee
In the conventions of \cite{Kachru:2002kx}, we have
\bena
&&g^1=\frac{e^1-e^3}{\sqrt{2}},
 \ \ \ \ \ g^2=\frac{e^2-e^4}{\sqrt{2}}, \nonumber \\
&&g^3=\frac{e^1+e^3}{\sqrt{2}},
\ \ \ \ \ g^4=\frac{e^2+e^4}{\sqrt{2}}\ ,\nonumber\\
&&g^5=e^5,  \nonumber
\eena
where, in terms of the angular coordinates $\psi$
in the range from 0 to $4\pi$ and $(\theta_1,\phi_1)$ and $(\theta_2,\phi_2)$ which parameterize two
$S^2$'s,
\bena
&&e^1=-\sin\theta_1 d\phi_1, ~~~~~~e^2=d\theta_1\ ,\nonumber \\
&&e^3=\cos \psi \sin\theta_2 d\phi_2-\sin\psi d\theta_2 \ ,\nonumber \\
&&e^4=\sin \psi \sin\theta_2 d\phi_2+\cos\psi d\theta_2 \ ,\\
&&e^5=d\psi+\cos \theta_1 d\phi_1+\cos \theta_2 d\phi_2 \nonumber \,
..
\eena
The warping factor $h$ is given by
\be h(\tau)= 2^{2/3} \mu^2\epsilon^{-8/3} I(\tau) \ ,\ee
where
\be
I(\tau)=\int_{\tau}^\infty dx \frac{x \coth x
-1}{\sinh^2x}\left(\sinh(2x)-2x\right)^{1/3}\ ,
\ee
and $\mu=2\pi^3 L^4 V_3 g_s l_s^2 \kappa_{10}^{-2}$.

The small and large $\tau$ regions in the KS background are easily
 found by recalling  that
\be
I(\tau)\sim .71805+{\cal{O}}(\tau^2)\, , ~~~~K(\tau)\sim \left(\frac{2}{3}\right)^{1/3}+{\cal{O}}(\tau^2)\, , && ~~~\mbox{ small $\tau$}\ ;\label{st}\\
I(\tau)\sim 3\times 2^{-1/3}\left(\tau-\frac{1}{4}\right)
e^{-\frac{4\tau}{3}}\, , ~~~~K(\tau)\sim 2^{1/3}
e^{-\frac{\tau}{3}}\, ,&&~~~ \mbox{ large $\tau$}\ .\label{lt}
\ee
The motion of the probe $D3$-brane in this background is described
by the DBI action. In compatibility with its equations of motion, we put the probe at a fixed value
for the angular coordinates $\theta_2,\phi_2,\psi$ and we also fix $\theta_1=\pi/2$. Moreover we choose the probe coordinates
to be $\eta=\xi^0,x^i=\xi^i$. The resulting degrees of freedom are therefore
$\tau=\tau(\eta)$ and $\phi_1=\phi_1(\eta)$. The induced metric is
\be
\!\!\!\!\!\!\!\!\!\!\!\!\!\!\!\!\!\! ds_i^2=-h^{-1/2}\left(1-\frac{\epsilon^{4/3} h}{6K(\tau)^2} \,
\tau'^2 - \frac{\epsilon^{4/3}h}{4}\cosh \tau K(\tau) \,
\phi_1'^2\right) d{\eta}^2+h^{-1/2} d\vec{x}\cdot d\vec{x}\ .
\ee
Then, the DBI action turns out to be
\ben
\!\!\!\!\!\!\!\!\!\!\!\!\!\!\!\!\!  \!\!\!\!\!\!\!\!\!\!\!\!\!\!\!\!\! S_{BI}=
-T_3V_3\int d\eta\left(\!\frac 1h\sqrt{1\!-\!\frac{\epsilon^{4/3}
h}{6K(\tau)^2} \,\tau'^2 - \frac{\epsilon^{4/3}h}{4}\cosh \tau
K(\tau) \,{\phi'}_1^2}\!-q\left(1-\frac 1h\right)\!\right)\ ,
\een
and, to leading order in the non-relativistic limit, we have
\be
S_{BI}=-T_3V_3\int d\eta \left(\frac{\epsilon^{4/3}}{12 K(\tau)^2 } \tau'^2+
\frac{\epsilon^{4/3}}{8}\cosh \tau K(\tau)\phi_1'^2\right)\ .
\ee
In terms of the brane cosmic time $t$ defined similarly to (\ref{proper}),
which in the small velocity limit turns out to be $dt/d\eta=h^{-1/4}$, the DBI action is written as
\be
S_{BI}=
-T_3V_3\int dt \,h^{-1/4}\!\left(\frac{\epsilon^{4/3}}{12 K(\tau)^2 }\,
\dot \tau^2+ \frac{\epsilon^{4/3}}{8}\cosh(\tau)\,  K(\tau)\,
\dot{\phi_1}^2\right)\ .
\ee
For this system, the conserved energy $U$ and angular momentum $l$ are
\begin{eqnarray}
U&=&\frac{\epsilon^{4/3}}{12 h^{1/4}K(\tau)^2}\left(
\dot \tau^2+\frac{3}{2} \cosh \tau\, K(\tau)\, \dot{\phi_1}^2\right)\ ,
\nonumber\\
l&=&\frac{\epsilon^{4/3}\sqrt\mu \cosh \tau\, K(\tau)\, \dot{\phi_1}}{4 h^{1/4}}\ .
\end{eqnarray}
The equations of motion for $\tau$ then read
\be
\dot \tau^2=\frac{12 h^{1/4}K(\tau)^2 U}{\epsilon^{4/3}}-\frac{24 h^{1/2}K(\tau) l^2/\mu}{\epsilon^{8/3}\cosh \tau}\ .
\ee

Recalling that the induced metric on the probe $D3$-brane
takes the FRW form with scale factor (\ref{scale}),
we get that the acceleration is
 \be
 \ddot a=\dot \tau\frac{\partial}{\partial \tau}\left(\dot \tau\frac{\partial h^{-1/4}}{\partial \tau}\right)\ .
 \ee
We should note that for small $\tau$ the mirage cosmology is very
simple, as
\be a(t)=a_0+{\cal{O}}(t^2)\, , ~~~a_0={\rm const.} \ee
due to Eq.(\ref{st}). For large $\tau$ on the other hand we get
that
\be
\!\!\!\!\!\!\!\!\!\!\!\!\!\!\!\!\!\!\!\!\!\!\!\!\!\!\!\!\!
\dot \tau^2=
4\cdot 2^{3/4}3^{5/4}\epsilon^{-2}\mu^{1/2}U\left(\tau-\frac{1}{4}\right)^{1/4}e^{-\tau}
-48\cdot 6^{1/2}\epsilon^{-4}l^2\mu \left(\tau-\frac{1}{4}\right)^{1/2} e^{-2\tau} \ ,
\ee
whereas for the acceleration we find
\be
\!\!\!\!\!\!\!\!\!\!\!\!\!\!\!\!\!\!\!\!\!\!\!\!\!\!\!\!\!\!\!\!\!\!\!
\ddot a= \frac{16 \cdot 2^{2/3}\epsilon^{-10/3}}{3(4\tau-1)}
\Big{[}-e^{-2\frac{\tau}{3}}\epsilon^{2}(2\tau^2\!-\!\tau\!-\!10)U+
\frac{2\cdot 6^{1/4} l^2}{\sqrt\mu} (4\tau-1)^{1/4}(8\tau^2\!-\!10\tau\!-\!7)e^{-5\frac{\tau}{3}}\Big{]}
\, .\nonumber\\\ee
We see that there are two contributions to the acceleration. The
first one is proportional to the energy $U$ and enters with a negative
contribution and so it leads to deceleration. The second
contribution is due to angular momentum $l$ and contributes positively so that it tends to accelerate the probe.
These terms are competitive and may lead to an acceleration period
as long as the angular momentum dominates the energy. However, at
the end energy always dominates (as it is multiplied by a factor
$e^{-2\tau/3}$ whereas the angular momentum is multiplied by a
factor $e^{-5\tau/3}$) leading to a final deceleration epoch generating a bounce in the probe brane trajectory. We would like to point out that the
nature of the bounce we describe here differs from the one found in \cite{Kachru:2002kx}. In our case, the bounce is due to a non-zero angular momentum of the probe brane whereas in \cite{Kachru:2002kx}, the bounce is due to the resolution of the conifold singularity of the CY.

Again one can show numerically that the numbers of e-folds during the acceleration period is too small to obtain inflation.

For large $\tau$, the KS metric after appropriate change of
coordinates, simplifies to the Klebanov-Tseytlin (KT) metric~\cite{Klebanov:2000nc}
\be
ds^2=\frac{r^2}{L^2\sqrt{\log(r/r_s)}}dx_{||}^2+\frac{L^2\sqrt{\log(r/r_s)}}{r^2}dr^2
+L^2\sqrt{\ln(r/r_s)}ds_{\mathbb{T}^{1,1}}^2\ , \ee
where
$ds_{\mathbb{T}^{1,1}}^2$ is the metric on the $\mathbb{T}^{1,1}$
manifold and $r_s=3^{1/2}2^{-5/6} \epsilon^{2/3}$.

\subsection{A red spectral index}
The non-relativistic limit of the DBI action in the KT geometry is
\bena
S&=&
-\frac{T_3}{2}
\int d^4x \left(
\partial_\mu r\partial^\mu r +
r^2\partial_\mu\phi_1\partial^\mu\phi_1
\right)\ . \label{dbikt}
\eena
In this case, in the $r\gg r_s$ ({\em i.e.} far form the singularity $r=r_s$) the Bardeen potentials are again
\be
\Phi= -\frac{\delta r}{r}\, , ~~~~ \Psi=-\Phi\ .
 \ee
From these last two equations, we deduce that the analysis of section \ref{perturbations} is valid and the power spectrum of
density perturbations is given as in Eq.(\ref{ppp}) by
\be
P= \frac{1}{2\pi^2}
 \frac{|\delta r_k|^2}{r^2}=\frac{1}{2\pi^2}\frac{|\delta r_k(\eta_*)|^2}{r_*^2}\ .
\ee

\begin{figure}[t]
\centering
\includegraphics[angle=0,width=4in]{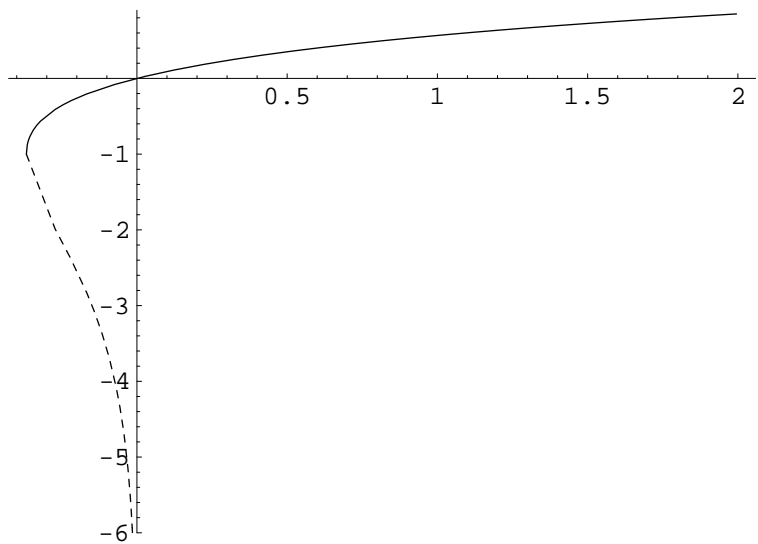}
\caption{ The two real branches of the Lambert W-function which satisfies $z=W(z) e^{W(z)}$. $W_0$ is the continuous
line and $W_{-1}$ is the dashed line.
\label{w}}
\end{figure}

In the present case, the scale factor is given by
\be
a_*=\frac{r_*}{L\big{(}\log(r_*/r_s)\big{)}^{1/4}}
\ee
and we should impose the  condition
 $a_*=k l_c$, which is explicitly written as
\be
\frac{r_*}{L\big{(}\log(r_*/r_s)\big{)}^{1/4}}=k l_c\ .
\ee
Defining $\zeta=({\sqrt{2} r_s}/{Lkl_c})^4$,
a real solution of the above equation turns out to be
\be
r_*=r_s \exp\!\left[\!\!-\frac{1}{4}W_{\!-1}\!\!\left(\!-\zeta\right)\right] ~~~~\mbox{for}
~~~~\zeta\leq e^{-1}\ ,
\ee
where $W_{-1}\!(x)$ is the second real branch of Lambert W-function, a plot of which is given in Fig.\ref{w}.
Thus, the power spectrum is given by
\be
P=2 \pi l_s^4
g_s  \frac{1}{k r_s^2} \, \exp\!\left[\!\frac{1}{2}W_{-1}\left(-\zeta\right)\right]\ ,
\label{pp}
\ee
whereas, by defining the scalar spectral index as $n_s-1=d\ln(k^3P)/d\ln k$, we find that
\be
n_s=\frac{3+W_{-1}\left(-\zeta\right)}{1+W_{-1}\left(-\zeta\right)}\ .
\ee
Recalling that~\cite{WW}
\be
W_{-1}(-x)=\ln(x)-\ln\big{(}-\ln(x)\big{)}+...\, ~~~~
\mbox{for}~~~~~x\to 0^{-}
\ee
we get for $\sqrt{2} r_s\ll L k l_c$
the following power spectrum
\be
P \simeq \frac1{k^3} \sqrt{\frac{\pi g_s}{N}} \frac{l_s^2}{ l_c^2}\, \frac1{\sqrt{{\ln(\frac{L kl_c}{\sqrt{2} r_s})}}}\ ,
\label{pppo}
\ee
and scalar spectral index
\be
n_s\simeq 1-\frac{1}{2\ln\left(\frac{L kl_c}{\sqrt{2} r_s}\right)}\ .\label{redd}
\ee
Considering the pivot wave number $k_p\approx 0.002\ \mbox{Mpc}^{-1}\ a_0$, where $a_0$ is the scale factor today,
we find that, in order to obtain the WMAP scalar spectral index at the pivot scale ($n_s\approx .95$), we need
\be
r_s\approx 6\times 10^{-8}\mbox{Mpc}^{-1}\ L\, l_c\, a_0\ ,
\ee
which gives a measure for the radius of the blown up sphere at the tip of the CY. Using this value in the expression (\ref{pppo}) for the power spectrum we get
\be
Pk^3 = 0.5 \times   \sqrt{\frac{g_s}{N}}\frac{l_s^2}{l_c^2} \simeq 10^{-10}
\ee
which is a constraint in our parameters.

\section{Collapsing length and compatibility of constraints}

{\noindent \bf{Collapsing length}} The first important point
to address is the definition of our collapsing length $l_c$.
General arguments imply the following string space-time
uncer\-tainty relation \cite{Li:1996rp}
\be
\Delta X\Delta t\gtrsim l_s^2\ .
\label{first}
\ee
One can show that the smallest scale that can be probed is the $11$-dimensional M-Theory
Planck scale \cite{Li:1996rp}, which is $l_P^{11}\sim g_s^{1/3}l_s$.
With that $\Delta X>g_s^{1/3}l_s$.
For a single perturbative mode the period $\Delta t$ of oscillation
is the wave length ($\lambda$). From (\ref{first}) we then have
\be
\lambda>l_s g_s^{-1/3}\ .\label{second}
\ee
This inequality implies a bound in the smallest wavelength that can be
probed semiclassically. Following this argument, our collapsing
length $l_c$ cannot be shorter than $l_s g_s^{-1/3}$. Since the only
fundamental parameters appearing in our model are $l_s$ and $g_s$ we
can therefore write
\be
l_c=l_s g_s^{-\gamma}\ ,
\label{lf}
\ee
where $\gamma>1/3$ is a constant.

An example of a possible value for $\gamma$ is the following. The first massive mode in the quantization of strings has a mass
$M\sim l_s g_s^{-1}$. Therefore one might decide that $M$ naturally sets the scale of
which processes can be treated semiclassically. Moreover, as $g_s\ll 1$ we have $l_s g_s^{-1}> l_s g_s^{-1/3}$ so that,
considering (\ref{second}) we might identify $l_c\sim l_s g_s^{-1}$.

From another perspective it is known that $D$-brane can probe scales of order $g_s l_s$
\cite{Shenker:1995xq}.
In other words we can say that the $D$-brane is only approximately thin, but instead has a
thickness of order $g_s l_s$. Using again the string space-time uncertainty relation we get $l_c\sim g_s^{-1} l_s$.
It is therefore reasonable for us to use $\gamma=1$ although we will keep this number undefined by the time being.

\vspace{.2cm}
{\noindent \bf{Constraints}} There is a number of potentially
controversial issues of the present setup that we would like to
discuss.

\begin{enumerate}

\item The first one regards the choice of physical frame.

To fix his units of length, a brane observer may use one of the dimensionful
parameters: $G_N$ (the four dimensional Newton constant), or
$m_{0}$ (the mass of a given elementary particle).
According to \cite{Grinstein:2000ny},
if the observer uses proper distance to measure
lengths on the brane, masses are necessarily fixed.
Following the lines of \cite{Kachru:2002kx}, we therefore consider as a physical frame the one where proper distances are used.

In the Slingshot Scenario, a warped throat is glued into a compact CY space.
At early times, when our wandering brane is on the throat,
ten dimensional graviton wave function has a non-trivial dependence on
the radial direction. This results in a time dependent Newtonian
constant of the effective four dimensional theory as seen for an
observer in the brane. In the adiabatic
approximation, in $AdS_5\times S_5$, $G_N\sim  a(t)^2/a_0^2\ M_p^{-2}$ \cite{Kachru:2002kx}, with $a_0$ the present value of the scale
factor\footnote{Here there is a possible problem of terminology. What we called mirage matter in the right hand side of eq. (3.1) can be re-interpreted as $\frac{8\pi G}{3}\rho_{\mbox{\tiny mirage}}$ where $\rho_{\mbox{\tiny mirage}}$ is the effective mirage energy density. In this way we can directly compare the mirage energy densities with local matter energy densities.}.
However, when at late times the brane reaches the hat of the CY, local gravity with a fixed Newtonian constant,
might be realized \cite{CottaRamusino:2006iu}.

Our model therefore results in a early-time variable $G_N$ cosmology, smoothly
joined into a constant $G_N$ late-time cosmology, {\it i.e.} the ``Brans-Dicke'' frame of the early times can be smoothly joined to an ``Einstein''
frame at late times.

Note that the observable consequences of the mirage era, where
$G_N$ is time dependent, have been proved to agree with
observations along the previous sections (flatness, isotropy,
power spectrum). Constraints on the time variation of $G_N$ become
relevant only at late times, where local gravity dominates. During local gravity era however, $G_N$ is constant and standard results follows.

An extra point is about the absence of singularity in our frame. One may argue that a conformal transformation to
the (early times un-physical) Einstein frame can bring back the cosmological singularity. However, as the change to the Einstein frame involves a
a smooth non-singular transformation, an early time Einstein frame is non singular as well.

\item
The second point
concerns the generation of the density perturbations as has been
criticized in~\cite{Kofman:2002cj}. In the model of \cite{Hollands:2002yb} the density
perturbations are generated at a fundamental scale $l_f$ with
$l_f>H^{-1}$. The fact that the fundamental scale $l_f$ is bigger
that the horizon has been emphasized in~\cite{Kofman:2002cj} as it sounds
quite unnatural.

In our case the fundamental length is not $l_c$ but the string length $l_s$ and one can easily check that
this length is the smallest length in our model.

Although for some parameters it can happen that $l_c>H^{-1}$ this
should not be seen as a problem in our model. In fact, $H^{-1}$ has
a physical meaning only as a measure of causality for a
four-dimensional observer. From the ten-dimensional point of view
the semiclassical approximation is under control as the brane orbit
is macroscopical. Instead, $l_c$ define the microscopical structure
of the probe brane.

\item Another point we would like to mention  appeared also in~\cite{Kofman:2002cj} and it can be formulated
in our case as follows. The scale factor $a_{*_{hor}}$ where the fluctuations, corresponding to the
present horizon of $10^{28} \rm{cm}$, were frozen, have the value
\be
a_{*_{hor}}=\frac{l_c}{10^{28} \rm{cm}} a_0\label{as} \ ,
\ee
with $a_0$  the present value of the scale factor. Note that $a_0$ cannot be set to one as usual,
as we have already chosen $a_b$ at the bouncing point. With a  radiation density today $\rho_r
\simeq 10^{-35}\rm{gr/cm}^3$, we get that the radiation density at the moment $\eta_{*_{hor}}$ was
\be
\rho_{*_{hor}}\simeq \rho_0 \left(\frac{a_0}{a_{*_{hor}}}\right)^{4}=10^{-17} M_p^4 \big{(}\frac{\mbox{cm}}{l_c}\big{)}^4\, .
\ee
Demanding that the four dimensional curvature is small, so as to keep the validity
of classical gravity, we should have  $G_N \rho_{*_{hor}}\ll M_p^2$, however as $G_N\sim  a(t)^2/a_0^2\ M_p^{-2}$ \cite{Kachru:2002kx},
we get that
\be
G_N\rho_{*_{hor}}\simeq 10^{-73} a_0^{-2}\ M_p^2 \big{(}\frac{\mbox{cm}}{l_c}\big{)}^2\ ,
\ee
and thus we have indeed $G_N \rho_{*_{hor}}\ll M_p^2$ for
$
l_c\gg 10^{-36} a_0^{-2}{\rm cm}.
$
Since $a_0> 1$, a sufficient condition is $\label{ls}
l_c\gg 10^{-36} a_0^{-2}{\rm cm}.$
\item \label{horizon} If this mechanism should be applied to the observed power spectrum,
an obvious constraint is that all density perturbations that crossed the horizon
from the CMB ($\sim 10^{24}$ cm) to the present cosmological
horizon of $10^{28}\mbox{cm}$ should be born after the bounce. A sufficient condition
is therefore $a_{*_{hor}}>a_b$ or
\be
\label{const1}
\frac{l_c}{10^{28}\cm}{a_0}
>\frac{l}{\sqrt{2U}}\ .
\ee
As we have noted above, $a_0$ cannot be fixed to one but it is rather a parameter and
(\ref{const1}) is a constraint on the value of $a_0$.

\item \label{frozen} A further constraint is obtained by the requirement that the relevant modes for density perturbations are
frozen immediately after they are created {\em i.e.}
\be
k_*<\frac{lL}{r_*^2}\ .
\ee

Moreover, they have to remain frozen at least until the local
gravity era.  If we choose an arbitrary matching point $r_{Match}$
as the end of the mirage era, we can see that a sufficient condition for that
is
\be
k_{CMB}<\frac{lL}{r_{Match}^2}\ ,
\ee
where $k_{CMB}$ is the
wave-number of the mode crossing the horizon at the CMB.

\end{enumerate}
\noindent There is also  a number of additional constraints on the parameters appeared
all along the previous sections whose compatibility should also be checked.

\begin{enumerate}
\setcounter{enumi}{5}
\item The degrees of freedom of the probe brane where described in terms of its embedding fields, whose dynamics
is controlled by the DBI action (\ref{actionn}). This in turn implies that the probe brane never
approach the stack, so that the open string modes related to the strings extending between them are always massive
$r_b\gg l_s$.
\label{born}
\item
\label{flatness}
In section \ref{early} we found that, in order to solve the flatness problem,  we need to enforce the condition
\be
l\ll  10^{-4}U\ .
\label{ineq2}
\ee
\item
\label{NonRel}
The calculations of section \ref{perturbations} are valid when the non-relativistic approximation
is satisfied
\be
{l^4}\gg 8{U^3}\ .
\label{nonrel2}
\ee
\end{enumerate}
\vspace{.1cm}

{\noindent \bf{Cross checking the constraints}}
Recalling that $k_* = a_*/l_c$, we can rewrite condition \ref{frozen} as $a_*^3 <  l\, l_c/L$. Now using the
evident relation $a_b<a_*$ we obtain
\be
{l^4}<  \big{(}\frac {l_c}L\big{)}^2 8 U^{3}\ .
\ee
In order to make this compatible with constraint \ref{NonRel} we need the following necessary condition
\be
\frac{l_c}{L}\gg 1\ .
\label{qqq}
\ee
Substituting in (\ref{qqq}) the definition of $L$ (\ref{L}) we obtain
\be
g_s^{-4\gamma}\gg Ng_s\ .\label{easy}
\ee
The constraint (\ref{easy}) can  easily be satisfied in the weakly coupled type IIB string theory
($g_s\ll 1$) and it can therefore be compatible with supergravity approximation ($g_s N\gg 1$) for positive $\gamma$.

\begin{figure}[t]
\centering
\includegraphics[angle=0,width=5in]{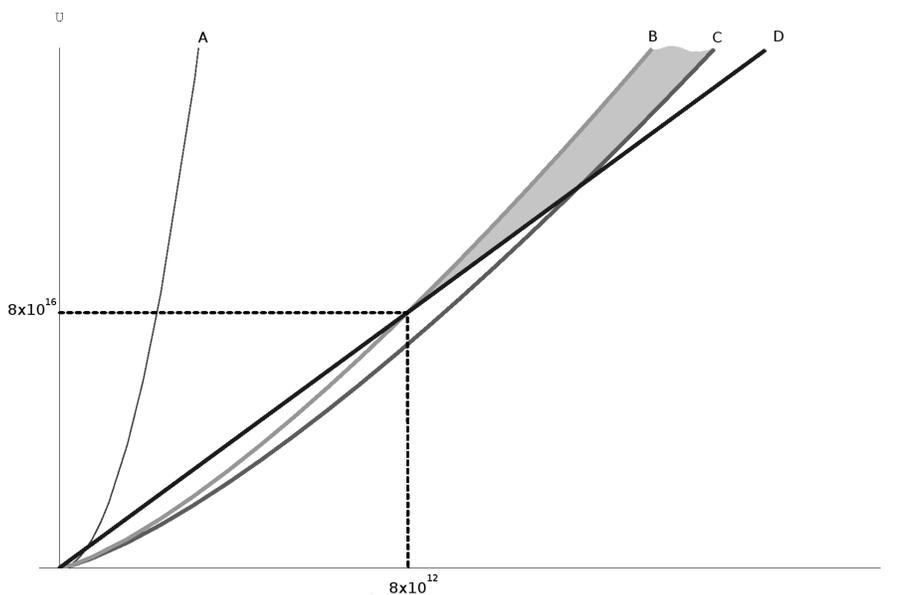}
\caption{The angular momentum and energy $(l,U)$ parameter space.
The region in which our model predicts observed results is marked on
gray. It is given by the intersection of the regions bellow lines A
and B and above lines C and D. Line A represents the region of
validity of the DBI action, Eq.(5.16) and it moves to the $U$ axis
as $g_sN$ grows. Line B is the limit of the non-relativistic
approximation Eq.(5.11), and it coincides with the frontier of the
region in which we have a bounce Eq.(2.22). Line D represents the
condition in order to solve the flatness problem Eq.(5.10). Finally
the constraint Eq (5.12) is represented by line C. We observe that
the width of the allowed region grows as $l^{4/3}$, meaning that as
we enlarge the angular momentum we relax the amount of fine tuning
needed to get phenomenologically reliable results. }
\label{parameters}
\end{figure}

The conditions \ref{flatness} and \ref{NonRel} constrain only the
energy $U$ and the angular momentum $l$. The non-relativistic approximation (\ref{nonrel2}) is compatible with
the small spatial curvature constraint (\ref{ineq2}) in the dashed region of Fig.\ref{parameters}. Moreover
they implies the necessary condition
\be
\label{top}
U \gg 8 \times 10^{16} \ , \label{Ul}
\ee
and then (\ref{ineq2}) and (\ref{nonrel2}) can simultaneously be satisfied.

Condition \ref{born} is ensured for $r_b\gg l_s$,  which can be
written, using Eqs.(\ref{L}, \ref{rb}) as
\be
 \pi l^4g_sN\gg U^2\, .
\label{612}
\ee
For $g_sN$ large enough this is satisfied in the dashed region of Fig.\ref{parameters}. This
makes \ref{born} consistent with the rest.

The last point involves the observed amplitude of the power spectrum.
Let us consider the flat spectrum of $AdS_5\times S^5$ (\ref{flat}).
Recalling the definition (\ref{L}) and (\ref{lf}), considering the fact
that the observed amplitude of the power spectrum is of order $10^{-10}$, we find
the constraint
\be
g_sN\sim 10^{19} g_s^{4\gamma+2}
\label{unca}
\ee
To make it compatible with the supergravity approximation $g_s\ll 1$, $g_s N\gg 1 \label{gs}$
we need
\be
10^{-\frac{19}{4\gamma+2}}\ll g_s\ll 1\ ,
\label{mag}
\ee
which is obviously satisfied for any $\gamma>0$ in accordance with previous discussions.

Thus, after all these constraints, it is interesting that we are not left with an empty set in the $(l,U)$ parameter space as it
is illustrated in Fig.3. The allowed region of parameter space is large enough, and it
requires no  fine tuning to achieve phenomenologically relevant results.

\section{Summary and Conclusions}

In this paper we have proposed a cosmological
model within the String Theory framework, where most of the
standard cosmological problems are naturally solved in the probe brane approximation.

In our model, the observable universe is a $D3$-brane moving in a
warped throat in a CY compactification of IIB string theory where we suppose that early
cosmology is dominated by a mirage era, while local
gravity becomes important only at late times when the probe brane
universe leaves the throat entering into the CY.

\begin{figure}[t]
\centering
\includegraphics[angle=0,width=4.5in]{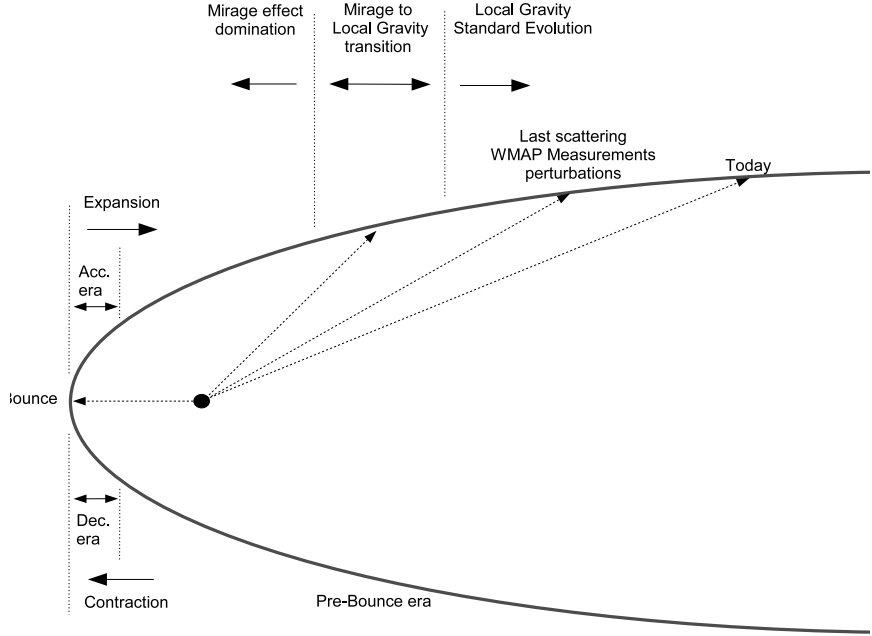}
\vskip-2.3in \caption{The cosmological  Slingshot model: the probe
brane follows an open orbit in the throat background. It comes
from a pre-bounce contracting past in the lower branch of the
orbit, reaches the turning point in which we have a cosmological
bounce, and then re-expands with a short acceleration period. The
early expansion epoch is dominated by mirage terms. At some point
we match it with standard cosmology and local gravity becomes
relevant, while the brane leaves the throat hitting the CY.
\label{model}}
\end{figure}

In this scenario, considering a non-vanishing value of angular momentum for the brane trajectory in
transverse space, we show the existence of a turning point on the
brane orbit at a finite distance of the tip of the throat.
From the point of view of an observer living in the brane, the
turning point prevents an initial singularity and gives rise to a
bouncing cosmology without passing through a quantum gravity
regime. There is a decelerating phase followed by an accelerating
one around the bounce, with a small e-folds number. The main
features of the model are depicted in Fig.\ref{model} and its achievements can be summarized as follows

\begin{enumerate}
\item {\bf Horizon} The well defined classical bounce (see section \ref{wandering})
solves the horizon problem of standard cosmology.
In fact, since there is no initial singularity the comoving horizon becomes infinitely bigger than the present Hubble
horizon, see Eq.(\ref{comoving}). In other words there have been
time enough for any two points in space to enter in causal
contact in the past.
\item {\bf Isotropy} As can be seen in equation (\ref{h2}),
at high energies the mirage energy contribution to the Hubble
equation goes like $a^{-8}$ so to dominate against anisotropic
perturbations (c.f. \ref{shear}), as shear goes like $\sigma^2\sim
a^{-6}$. Shear instability is therefore not present in the
Slingshot scenario, solving the isotropy problem of standard
cosmology.
\item {\bf Flatness} Regarding spatial flatness, we calculated the deviation
$|\Omega_k-1|$ from a flat three-dimensional spatial section in
Eqs.(\ref{omegaminusone})-(\ref{37}) showing that with a non-vanishing angular momentum, it is bounded from below.
The minimum of $|\Omega_k-1|$ is a function the orbital parameters
of the brane trajectory $l,U$. Demanding the spatial curvature to be small imposes only
one inequality (see (\ref{ornitorrinco})) on the two-dimensional
parameter space, leaving half of the space at our disposal,
(c.f. Eq.(\ref{ineq})). In conclusion, we get a spatially flat universe with a
very generic choice of parameters, no fine-tuning is needed. The
flatness problem is then solved.
\item {\bf Spectral Index} To calculate the power spectrum of density perturbations, we
assumed in section \ref{perturbations} that there is a fundamental length $l_c$ at which a
quantum fluctuation collapses and the corresponding perturbation
behaves classically. The value of $l_c$ can be inferred from the
String Spacetime Uncertainty Relation (\ref{first}), getting
(\ref{lf}). After such collapse has occurred, the existence of
non-zero angular momentum allows for frozen modes at large scales
to appear. These modes survive up to late
time, giving rise to a slightly red power spectrum, as can be seen
Eq.(\ref{redd}). Also we argued that tensor perturbations are suppressed.

\end{enumerate}

The main predictive achievement of our Slingshot scenario is the calculation of the spectral index for density perturbations.
Although the mechanism we use to
produce these perturbations is similar to the one used by \cite{Hollands:2002yb}, we showed that
the reservations expressed in \cite{Kofman:2002cj}
are naturally solved in our model.

Apart from the above achievements the strength of our model resides also on its generality.
Brane inflationary scenarios indeed usually ignore angular-momentum contributions
\cite{Kachru:2003sx} (see also
\cite{Dvali:1998pa}-\cite{Matsuda:2006ie}). This is due to the fact
that in an expanding background any non vanishing angular momentum
is rapidly damped by the expansion. On the other hand, if the
background is static, zero angular-momentum is just a very special
choice in the space of all possible initial conditions for the
brane motion. In our case the absence of $\bar D3$-branes ensures
that we can consistently choose a static background, and then the
vanishing angular momentum might well be thought as a fine tuning. The
generic case of a nonzero impact parameter, {\em i.e.} a
non-vanishing value of the angular momentum, gives rise to a very
rich set of solutions \cite{Kehagias:1999vr, Kiritsis:2003mc, Burgess:2003qv, Brax:2002qw, Brax:2002fu}
and to the possibility of bouncing cosmologies \cite{Burgess:2003tz}. It is
indeed in this context, that our ``Cosmological Slingshot
Scenario'' is based.

There are many open issues of our model that is left for future research. An open problem for example
is to consider corrections to the probe brane approximation used here.
This is important  for the late time evolution when the brane leaves the throat of the CY. There,
according to our paradigm, the slingshot cosmology evolves into standard cosmology, or in other words, the mirage evolution is overcome by
local gravity.
This involves a transition from a mirage dominated era with a moving brane without any matter,
into a local gravity dominated era with an static brane and matter fields excited on it. This transition
has to be understood as an analogous of the reheating process in standard inflationary models. It entails
a dynamical mechanism under which the kinetic energy of the brane is passed to matter fields. The exact
description of this dynamics as well as the robustness of our predictions for physical observables is an
open point of the model which is left for future research.

\subsection*{\bf Note added}

After the first draft of this paper appeared on the web, some other papers were released \cite{Easson:2007fz}-\cite{Germani:2007uc}, whose
relation with the present work is worth to clarify.

In \cite{Easson:2007fz}, the orbits of a probe brane in a Klebanov-Strassler throat were
explored numerically using the full BDI equations of motion. The resulting mirage cosmology
corresponds to bouncing/cyclic universes. This confirms the presence of the bounce, studied
here in the nonrelativistic approximation of section \ref{KSsection},
and completes the analytic results obtained in the Klebanov-Tseytlin limit. It would be
interesting to know whether this results may improve the solution to the standard cosmological
problems presented in our paper.

In \cite{Easson:2007fz} the motion of an antibrane probe was studied as well. In this case, neglecting the
backreaction of the probe may be a catastrophic assumption: due to the nonvanishing vacuum energy
there is an exponential expansion of the 4d slice that gives rise to Hubble damping. Since the angular
momentum scales as $l^2\sim a^{-6}$, the corresponding centrifugal barrier disappears very fast, and the
system flows into a standard brane inflationary scenario \cite{Kachru:2003sx}. Nevertheless, based in
numerical calculations, in \cite{Easson:2007dh} was claimed that the presence of a nonvanishing angular
momentum may contribute with an additional few of e-folds to the inflationary era. These investigations
are not related to what we presented here, where we concentrated in the brane case.

Finally, in \cite{Brandenberger:2007by} the power spectrum for perturbations was calculated in a
complementary way, by making use of a different choice of vacuum and in the case of zero angular momentum.
Instead of assuming that perturbations are in their quantum mechanical vacuum when their wavelength is
smaller than a quantum critical length $l_c$, so that perturbations are continuously created in conformal
time, the authors in \cite{Brandenberger:2007by} assumed that the perturbations were created at the infinite
conformal past ({\it i.e.} well before the bounce). For this reason, and because the lack of angular
momentum in their case, the spectrum of primordial scalar perturbation
was found to be blue, in contrast to the slightly red spectral index found here. With the present knowledge of
the quantum mechanical behavior of branes at high energies and the infinite past evolution of the Slingshot Universe,
there is not yet a definitive answer to what it should be the "correct" vacuum. In \cite{Brandenberger:2007by} it is
was supposed that perturbations where quantum mechanically created at macroscopic scales (as at past infinity the
Universe is contracting). In the Slingshot instead the perturbations are created at some quantum mechanical scale defined
by the underline quantum String Theory.

\subsection*{Acknowledgments}
CG wish to thank Kate Marvel, Mairi Sakellariadou, David Wands
and Daniel Wesley for useful discussions on physical cosmology. He also thanks Cliff Burgess, Roy Maartens, Nik
Mavromatos and Shinji Tsujikawa for correspondence. CG~was supported by PPARC research grant PPA/P/S\-/2002/00208
during the early stages of this work. NEG wants to thank Samuel Leach and Paolo Creminelli for helpful discussions,
as well as Fernando Quevedo for help and encouragement in the early stages of this work. NEG is grateful to SISSA
and ICTP for hospitality and support during most of this work. This work is also supported by the RTN programme,
``Constituents, Fundamental Forces and Symmetries of the Universe", MRTN-CT-2004-005104.

\appendix

\section*{Appendix}

The existence of frozen decaying modes in the fluctuation spectrum can be shown by
different means directly in the $\delta r_k,\delta\Omega_k$ variables.
By using the background solution Eq.(\ref{8}) and after Fourier transforming, the
quadratic action for the fluctuations is written as
\bena
\!\!\!\!\!\!\!\!\!\!\!\!\!\!\!\!\!
S=\frac{T_3}{2}\int d\eta
\left(
\phantom{\frac12}\!\!\! {\delta r}_k'^2 + r^2\delta {\Omega}'^2_k -\left(k^2-\Omega_5'^2
\right)
\delta r_k^2 - r^2k^2\delta\Omega_k^2 + 4r\, \Omega_5'\,\delta {\Omega}_k'\delta r_k
\right)\ .
\eena
The equations of motion for  the perturbations turn out then
to be
\bena
\label{pert}
\frac{d~}{d\eta}\left(r^2\delta\Omega'_k+\frac{2lL}{r}\delta r_k\right)+r^2k^2\delta\Omega_k=0\ ,
\label{p1}
\\
\delta r''_k  + \left(k^2-\frac{l^2L^2}{r^4}\right)\delta r_k - 2r\,\Omega_5'\,\delta {\Omega}'_k =0
\label{p2}\ ,
\eena
where Eq.(\ref{3}) has been used. A first integral of these equations can be found as
the energy associated to the above action
\be
\delta e_k =
\frac {T_3}2 \left(\phantom{\frac12}\!\!\! {\delta r'}_k^2 + r^2{\delta \Omega'}_k^2 +\left(k^2-\frac{l^2L^2}{r^4}
\right)\delta r_k^2
+r^2k^2\delta\Omega_k^2 \right)\ ,
\ee
so that the corresponding  potential is
\be
{\cal V}=\frac {T_3}2\left(\left(k^2-\frac{l^2L^2}{r^4}\right)\delta r_k^2 + r^2k^2\delta\Omega_k^2\right)\ .
\ee
For $k>lL/r^2= k_0$ the system clearly oscillates. However, $k<k_0$ (for
sufficiently small $\delta\Omega_k$),  we will see that the $\delta r_k$ perturbations grow immediately after they are created. The system
of equations although complicated enough, it can easily be solved for
$r\delta\Omega_k\ll \delta r_k$. This is physically reasonable as one
would expect that the angular momentum is important only at the
inversion point of the trajectory. Mathematically one can check
that for sub- and super-critical modes (i.e. $k\gg k_0$ and $k\ll k_0$)
this approximation is valid.

\begin{figure}[t]
\centering
\includegraphics[angle=0,width=3in]{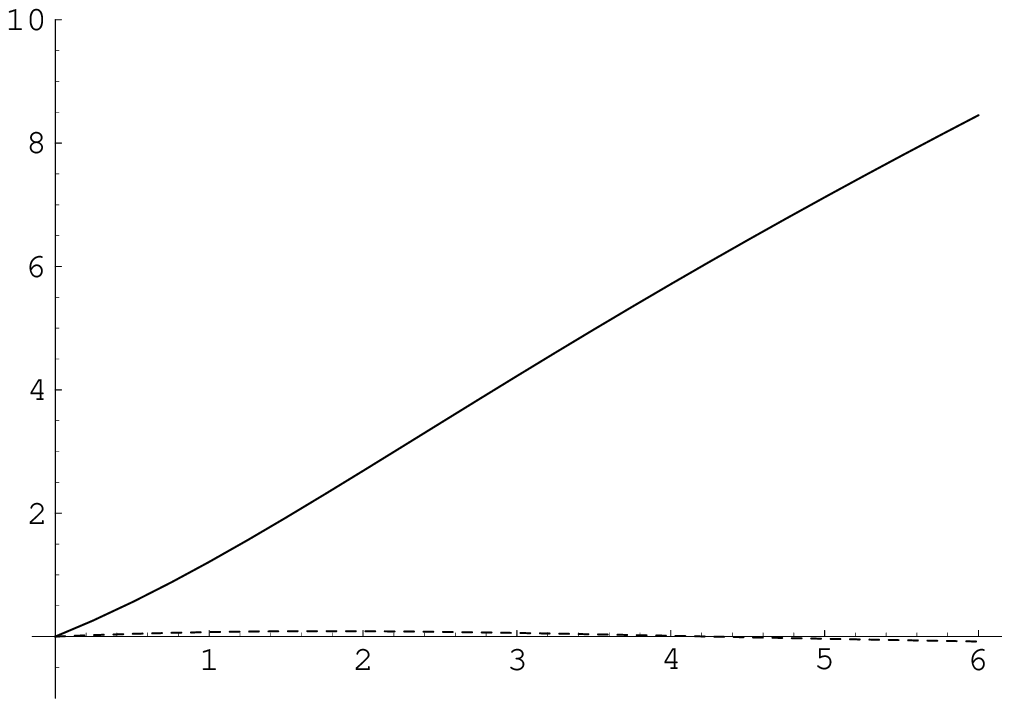}
\caption{ Numerical integration of Eqs.(5.14)
for $(k,l,E)=(0.002,25,4)$ and initial conditions $\delta r(0)=0,\delta r'(0)=1,\delta \Omega (0)=0, \delta\Omega'(0)=0.1$.
The continuous line is $\delta r(\eta)$ and $\delta\Omega(\eta)$ is the dashed line. Indeed, $\delta \Omega$ remains zero, while
$\delta r$ evolves linearly (as long as $k<2lL/r$).
\label{numerical}}
\end{figure}

For sub-critical modes substituting $\delta \Omega_k '$ from the
second equation (\ref{p2}) to the first, it is possible to see
that $\delta\Omega_k=0$ is a good approximation whenever
$k\gg k_0$. In the case of super-critical modes ($k\ll k_0$) Eq.(\ref{p2}) turns out to be
\be
{\delta r}_k''  -\frac{4 (E-1)^2 l^2L^2}{(4(E-1)^2\eta^2+l^2L^2)^2}\delta r_k  =0 \label{eqr}
\ee
and the corresponding solution is
\be
\delta r_k= C_k\,  r\, ,
\label{rr}
\ee
where $C_k$ a constant whose value is obtained by matching  the
solution for $\delta r_k/r$, at the critical value $k=k_0$ with the corresponding
canonically normalized oscillating sub-critical mode, obtaining $C_k =
1/\sqrt{2k}$. In this case, this solution, together with $\delta
\Omega_k=0$, is an exact solution of the system
(\ref{p1})-(\ref{p2}) at $k=0$ and an approximate solution at $k\ll k_0$. This approximate solutions are also verified numerically,
see Fig.\ref{numerical} .

\end{document}